\documentclass[12pt,letterpaper]{article}

\usepackage[utf8]{inputenc} 
\usepackage[T1]{fontenc}    
\usepackage{times}          
\usepackage{amsmath}        
\usepackage{setspace}
\newcommand{\defaultspacing}{\setstretch{1}}
\usepackage{amsfonts}       
\usepackage{amssymb}        
\usepackage{graphicx}       
\usepackage{hyperref}       
\usepackage{natbib}         
\usepackage{url}            
\usepackage{booktabs}       
\usepackage{lipsum}         
\usepackage{algorithm}
\usepackage{algorithmic}
\usepackage{soul}
\usepackage{color}
\usepackage{newfloat}
\usepackage{listings}
\usepackage{float}
\usepackage{listings}
\usepackage{comment}
\usepackage{graphicx}
\usepackage{subcaption}
\usepackage{longtable}
\usepackage{caption}
\usepackage[table]{xcolor}

\usepackage{xr-hyper}        

\usepackage{xr}
\newcommand{\ImModel}{ViT}

\usepackage{tikz}
\usetikzlibrary{bayesnet}
\usetikzlibrary{arrows}
\usepackage{color}
\usetikzlibrary{backgrounds}
\usepackage{graphicx}
\newcommand{\SpatialRes}{6.7}

\usetikzlibrary{arrows.meta,positioning,fit}
\tikzset{
  obs/.style   = {circle, draw, thick, inner sep=1pt,
                  minimum size=17pt, font=\scriptsize},
  unobs/.style = {obs, densely dotted},
  arr/.style   = {-{Stealth[length=3.5pt]}, thick},
  title/.style = {font=\footnotesize, anchor=south}
}

\usepackage{amsmath}
\usepackage{amsfonts}
\usepackage{amssymb}

\usepackage{listings}

\newcommand{\RobustNCellsCommon}{25}

\newcommand{\RobustNTLMaxAbsDiff}{7.38}
\newcommand{\RobustNTLMedianAbsDiff}{2.05}

\newcommand{\RobustNTLCorrATE}{0.85}
\newcommand{\RobustLargeBuffCorrATE}{0.60}
\newcommand{\RobustStrictPrecCorrATE}{0.61}

\newcommand{\CrossSectorCanonicalXSalience}{0.182}

\newcommand{\WBMaxSector}{Trade and Tourism (330)}
\newcommand{\WBMaxSectorATE}{12.29}
\newcommand{\WBMinSector}{Government and Civil Society (150)}
\newcommand{\WBMinSectorATE}{-0.16}
\newcommand{\CHMaxSector}{Emergency Response (700)}
\newcommand{\CHMaxSectorATE}{15.15}
\newcommand{\CHMinSector}{Agriculture, Forestry and Fishing (310)}
\newcommand{\CHMinSectorATE}{1.21}

\newcommand{\AdjRsqModelOne}{0.73}

\newcommand{\AdjRsqModelTwo}{0.71}

\newcommand{\WBMinSectorUnitFEATE}{-0.05}
\newcommand{\WBMinSectorUnitFEATESE}{0.11}

\newcommand{\GlobalNUnits}{9,899}
\newcommand{\GlobalNCountries}{36}

\usepackage[margin=1in]{geometry}

\usepackage{authblk}

\title{
Chinese vs. World Bank Development Projects: 
\\ Insights from Earth Observation and Computer Vision on Wealth Gains in Africa, 2002-2013
}

\date{\today} 

\title{
Chinese vs. World Bank Development Projects: 
\\ Insights from Earth Observation and Computer Vision on Wealth Gains in Africa, 2002-2013
}

\author[1, 2, 4]{Adel Daoud\thanks{Corresponding author: adel.daoud@liu.se. All authors contributed equally. Forthcoming in \textit{World Development}.}}
\author[2, 4]{Cindy Conlin}
\author[3, 4]{Connor T. Jerzak}

\affil[1]{Chalmers University, Sweden \qquad $^2$Linköping University, Sweden}
\affil[3]{University of Texas at Austin, USA \qquad $^4$AI and Global Development Lab}

\begin{document}

\maketitle

\begin{abstract}
\noindent Debates about whether development projects improve living conditions persist, partly because observational estimates can be biased by incomplete adjustment and because reliable outcome data are scarce at the neighborhood level. We address both issues in a continent‑scale, sector‑specific evaluation of Chinese and World Bank projects across \GlobalNUnits{} neighborhoods in \GlobalNCountries{} African countries (2002-2013), representative of $\sim$88\% of the population. First, we use a recent dataset that measures living conditions with a machine‑learned wealth index derived from contemporaneous satellite imagery, yielding a consistent panel of 6.7 km square mosaics. Second, to strengthen identification, we proxy officials' map‑based placement criteria using pre‑treatment daytime satellite images and fuse these with tabular covariates to estimate funder‑ and sector‑specific ATEs via inverse‑probability weighting. Incorporating imagery often shrinks effects relative to tabular‑only models.
On average, both donors raise wealth, with larger and more consistent gains for China; sector extremes in our sample include \textit{\WBMaxSector{}} for the World Bank (+\WBMaxSectorATE{} IWI points), and \textit{\CHMaxSector{}} for China (+\CHMaxSectorATE{}). Assignment‑mechanism analyses also show World Bank placement is often more predictable from imagery alone (as well as from tabular covariates). This suggests that Chinese project placements are more driven by non-visible, political, or event-driven factors than World Bank placements. To probe residual concerns about selection on observables, we also estimate within‑neighborhood (unit) fixed‑effects models at a spatial resolution about 67 times finer than prior fixed-effects analyses, leveraging the computer-vision-imputed IWI panels; these deliver smaller but, for Chinese projects, directionally consistent effects.
Methodologically, we extend recent EO–ML causal inference frameworks by fusing pre‑treatment satellite imagery with tabular covariates to estimate treatment propensities, and by systematically benchmarking image‑augmented estimators against tabular‑only and unit fixed‑effects designs using new assignment‑mechanism diagnostics. Empirically, we provide a continent‑wide, sector‑specific comparison of the neighborhood‑level wealth effects of Chinese and World Bank projects across \GlobalNUnits{} African neighborhoods.
%
\end{abstract}

\newpage 

\defaultspacing

\section*{Introduction}\label{s:Intro}

The United Nations' establishment of the Millennium and Sustainable Development Goals, along with China's Belt and Road initiative, has sparked growing interest since 2000 in examining how aid and development programs help nations overcome poverty. Indeed, around 2005, China switched from being a net aid recipient to a net donor \citep{Chin2012}.  Scholars estimate China now spends twice as much as the U.S. and other major donors, like the World Bank \citep{Malik2021,daoudIMFFairnessCalibrating2022}, generating considerable interest among international actors about China’s increasing influence in developing nations, particularly Africa, and about how Chinese activities ``affect social, economic, environmental, and governance outcomes in [these] low- and middle-income countries'' \citep[p.~7]{Dreher2022}.  Between 2011 and 2020, an average of 82 papers per year examined aid effectiveness \citep{Asatullaeva2021}.

However, this large and expanding literature has not reached consensus on whether development programs are effective at lifting the living conditions of African people, producing economic growth and institutional development in recipient countries \citep{Ahmed2022,Asatullaeva2021,Mandon2023,McGillivray2006,hallerod_bad_2013,daoud_impact_2017,daoudImpactAusterityChildren2024}.  One way to improve the evidence base of development-programs efficiency is to conduct randomized controlled trials \citep{Banerjee2015,Duflo2015}, but these are challenging to conduct at the scale of the African continent. Thus, most aid studies are observational and face significant methodological challenges in identifying unbiased causal estimates that apply to the target population. 

There are two factors limiting such large-scale analyses.  First, limited capacity in recipient countries often results in missing or low-quality outcome data, especially at the sub-national level \cite{Daoud2022,burkeUsingSatelliteImagery2021}.  Such data would cost millions of USD to collect. With the potential dismantling of USAID and, consequently, the Demographic and Health Surveys (DHS), alternative methods are needed to collect data on people's health and living conditions in Africa and beyond.

Second, this limited capacity also hampers the availability of geo-temporal control variables that would enlarge the adjustment set for causal identification \cite{jerzak2023image}. For example, one such covariate set would be to access the maps that aid and development program officials used when allocating their programs. For example, China considers its foreign aid and loan activities a state secret. Official data on these activities is not publicly available, so studies of Chinese aid use a dataset collected from media reports, recipient governments, researchers, non-governmental organizations, and Chinese embassy websites \citep{Strange2017}.

This study contributes to the literature by applying two innovations that address data limitations and methodological challenges of evaluating the effectiveness of foreign aid from two distinct aid donors: China and the World Bank. We study the aid they provided to \GlobalNCountries{} African countries (representing 88\% of the continent’s population) between 2002 and 2013, based on a representative sample of \GlobalNUnits{} neighborhoods as the pool of potential treatment and control units of analysis.  

The first innovation of our study is that we use a new data source for subnational wealth at a resolution of 6.7 km over all African human settlement areas, generated by \citet{Pettersson2023}, who trained a computer vision machine learning algorithm to learn the relationship between daytime and nighttime satellite images and on-the-ground wealth measures from DHS from the same times and places.  Once the algorithm learned the association between what it saw in satellite images and wealth measures, it imputed wealth for areas without DHS wealth measures to create a continent-wide IWI measure for the entire African continent, from 1990 to 2020. This would not have been possible using DHS data alone, since surveys were not done repeatedly for the same neighborhoods and are available only in limited timeframes for each country; see Table \ref{tab:iwi_na_side_by_side}.

The second innovation of our article is that we apply a causal inference method, based on the planetary-causal-inference paradigm \cite{Jerzak2023,Jerzak2023a}, synthesizing \textit{causal} and \textit{predictive} logics \cite{daoud_statistical_2023}. This paradigm aims to advance research on global development by developing earth observation (EO) and machine learning (ML) methods for causal analysis and measurement---denoted \textit{EO-ML methods} as a shorthand. To enhance our identification approach, we use daytime satellite imagery over treated and control communities to adjust for factors visible to remote sensors (such as physical geography, topography, temperature, and infrastructure) and their spatial arrangement. These images will likely enhance identification because they provide additional information on why Chinese and World Bank officials will place their programs in some communities rather than others \cite{bediMorePrettyPicture2007}. Decision makers may look for specific patterns in their maps—proxied by our satellite images—when making sub-national project site placement decisions. These image-based confounders identify features that are hard to encode in tabular data.  

We provide our EO-ML causal method with daytime satellite images centered over treated and control communities, one period before the aid project commitment, along with a large set of tabular-format variables representing political, economic, and other factors that empirical research has shown are relevant to aid allocation and effectiveness. We employ EO-ML methods to identify features visible in satellite imagery (along with tabular covariates) to estimate each community’s treatment propensity. These propensities are used in an inverse probability weighting procedure to estimate the effect of aid from each funder and sector on neighborhood wealth. 

This EO-ML approach targets the average treatment effect (ATE) estimand for \GlobalNCountries{} African countries over the period 2002 to 2013. Many policy evaluation studies on Chinese and World Bank programs target the local average treatment effect (LATE) estimand, because they use an instrumental variable approach. Although LATE is considered to provide a more credible identification approach, this estimand may struggle with external validity because the variation it targets is much narrower than the ATE, as this variation relies on the compliers in the instrument. As our goal is to estimate an Africa-wide effect over the given time period, we target the ATE. Our expectation is that we will improve our identification by relying on an additional adjustment set—satellite images to capture selection-relevant covariates. 

Following recent contributions \citep{chai2023world}, our study analyzes aid by sector (e.g., \textit{Health}, \textit{Education}, \textit{Energy Generation \& Supply}), which is important due to the heterogeneity in placement logics, political economy factors, expected timeframes for results, and expected geographic reach across sectors.  
Earlier studies analyzed sector groups, such as early- vs. late-impact aid \citep{Dreher2015a} or Economic, Social, and Production groups \citep{Bitzer2018,Xu2020,Dreher2021b}. In contrast, our study analyzes individual aid sectors within the same framework; this enhances the comparability of aid across funders who prioritize different sectors.

The remainder of the paper proceeds as follows.  The next section examines the factors that influence the placement and effectiveness of foreign aid, with a focus on the processes of development projects from China and the World Bank.  Following this is a detailed description of the data and methods, accompanied by the results, discussion, opportunities for future research, and conclusion. Three appendices---Data, Results, and Code---are included separately and provide more detailed information.  

\section{The Political Economy of Development Projects and Human Development}

While a large literature examines the political and economic determinants of development-project allocation, a parallel and often contentious debate centers on a more fundamental question: Does aid, or other forms of development projects, actually improve the living conditions of people in recipient countries? \citep{Asatullaeva2021, McGillivray2006}. Answering this question is complicated by the diverse strategies of donors and the micro-macro paradox that aid can be locally effective yet hard to detect in aggregate outcomes. This study focuses on Africa’s two most significant development financiers---China and the World Bank---whose contrasting approaches provide a critical lens for examining the impact of aid on poverty, wealth, and local economic activity. China has recently become the single largest financier of African infrastructure, while the World Bank remains one of the largest traditional donors with a broad development mandate. Their sharply different models of aid delivery offer a unique opportunity to assess whether and how foreign aid improves living conditions on the ground.

\subsection{The Impact of Chinese Development Finance on Living Conditions}

China’s emergence as a major development partner in Africa has been characterized by a focus on ``hard'' infrastructure projects---roads, railways, ports, and power plants \citep{brautigam2009dragon, Dreher2022}. The primary mechanism through which this form of aid is believed to impact living conditions is by stimulating local economic activity. Sub-national studies using satellite nightlights as a proxy for economic growth consistently find that Chinese-funded infrastructure projects generate significant localized economic booms \citep{Bluhm2020, Dreher2021b}. This reflects China’s strategy of ``connective financing,'' which reduces transportation costs, links remote areas to markets, and can spur economic agglomeration, creating measurable spillovers in the local economy \citep{Donaldson2016}.

However, whether these localized growth spurts translate into broad-based improvements in household welfare is less clear. Some micro-level studies find that Chinese aid projects are associated with increased household consumption, asset ownership, and employment, suggesting a direct positive impact on poverty reduction \citep{martorano2020chinese,Brazys2021, Leiderer2021}. Yet, the benefits may not be evenly distributed. The political economy of Chinese aid allocation---which can favor the birth regions of political leaders \citep{Dreher2019} and often bypasses local governments and civil society channels---may concentrate economic gains in specific locales or among politically connected elites \citep{Gehring2022, Lee2021}. Moreover, although Chinese projects create employment opportunities, debates continue over the extent to which they rely on local versus imported Chinese labor and the quality of jobs created for African workers \citep{Warmerdam2013}.

China also funds numerous projects in social sectors, but the impact of these projects on living conditions is less studied and potentially more mixed. China’s demand-driven, state-led model may enable the rapid construction of schools and clinics, but some research suggests that these projects may be less effective at improving long-term outcomes compared to those of traditional donors \citep{Martina2020}. The reason often cited is a focus on ``hardware'' over ``software''---Chinese projects tend to emphasize the provision of physical infrastructure without an equivalent investment in the staffing, training, and institutional capacity needed to deliver services effectively.

\subsection{The Impact of World Bank Development Projects on Living Conditions}
In contrast to China’s infrastructure-heavy portfolio, the World Bank pursues a much broader development mandate that includes poverty reduction, human capital development, and governance reform \citep{WorldBank2008}. Its theory of change relies on more indirect and long-term mechanisms, aiming to build sustainable local capacity by working through recipient government systems---a principle formalized in the Paris Declaration on Aid Effectiveness \citep{OECD2005}.

Consequently, the measured short-run impact of World Bank aid on local economic activity is often smaller and less immediate than that of Chinese aid. Studies using nightlights data, for example, have difficulty detecting any significant uptick in luminosity around World Bank project sites \citep{Bitzer2018, Dreher2015a}. However, this does not necessarily mean the Bank’s projects have a weaker impact on living conditions; rather, they often operate through channels that nightlights fail to capture. The World Bank allocates a significant portion of its portfolio to social sectors, where investments in health and education yield long-term gains in human capital \citep{Krueger2001}. Empirical evidence bears this out: providing safe water has been shown to directly improve health outcomes by reducing disease incidence and child mortality \citep{Duflo2015, Galiani2005}. Yet, working through government ministries can slow implementation and create opportunities for elite capture \citep{Platteau2004}, and the Bank's priorities are not immune to the influence of its most powerful shareholders \citep{Andersen2006, Kim2021, Martel2021}. These biases can blunt the poverty-reducing impact of projects, despite the Bank's explicit pro-poor mandate \citep{Briggs2017}.

\section{Two Key Gaps}

Two key challenges emerge from this literature. First is the problem of measuring neighborhood-level socioeconomic outcomes \cite{kino_scoping_2021,Jean2016,Daoud2022,sakamotoScopingReviewEarth2025}. Most subnational impact studies have relied on satellite nighttime lights, a proxy that is better suited to capturing the effects of large-scale infrastructure than improvements in household well-being. Our study addresses this gap by using a novel wealth index derived from machine learning and high-resolution satellite imagery \citep{Pettersson2023}, thereby providing a more holistic view of living conditions.

Second is the persistent challenge of causal identification. Development programs are not randomly assigned \cite{bediMorePrettyPicture2007}. While past studies have used instrumental variables or fixed-effects models, these approaches may not fully account for fine-grained confounders related to geography and the built environment. Our study adapts and applies an EO–ML causal adjustment approach to further enhance identification \citep{sakamotoScopingReviewEarth2025}: we use the satellite images themselves as control variables, which also enables new kinds of substantive insights to be gained about the assignment mechanism surrounding aid \citep{rubin1991practical,daoud_statistical_2023}---capturing what kinds of information do and do not predict aid decisions. 
By conditioning on satellite images---which proxy for the maps used by aid officials in project allocation \citep{bedi2007more}---we compare treated and untreated areas with greater sensitivity, thereby enhancing the credibility of our causal estimates and addressing limitations in prior literature \citep{Easterly2005}. In this way, we can more robustly identify causal impacts of aid interventions while controlling for and explicitly characterizing the funder targeting strategies.

\subsection{
Measuring Strategic Allocation with Satellite Images \& Computer Vision
}

To work towards closing these two gaps, we proceed as follows. We begin with identification: we formalize the project placement problem and show how pre-treatment satellite imagery can proxy the map-based information donors use when deciding where to locate projects, strengthening causal adjustment in an observational setting. We then turn to measurement, introducing the satellite-derived wealth panel that makes neighborhood-level evaluation feasible at continental scale.

We start by laying out the treatment-assignment problem and clarifying the role of imagery as a proxy for otherwise-unobserved spatial selection factors in observational studies of development programs. We will estimate the effects of World Bank and Chinese projects on neighborhood-level living conditions in Africa, drawing on the potential outcomes framework to define estimands and invoking graphical models to clarify assumptions. Our strategy leverages satellite imagery not merely as a covariate but as a proxy for the geospatial features that likely inform donors' assignment decisions, thereby clarifying aspects of the program allocation mechanism itself \cite{bediMorePrettyPicture2007}.

\begin{figure}[htb]
  \centering
\includegraphics[width=1.0\linewidth]{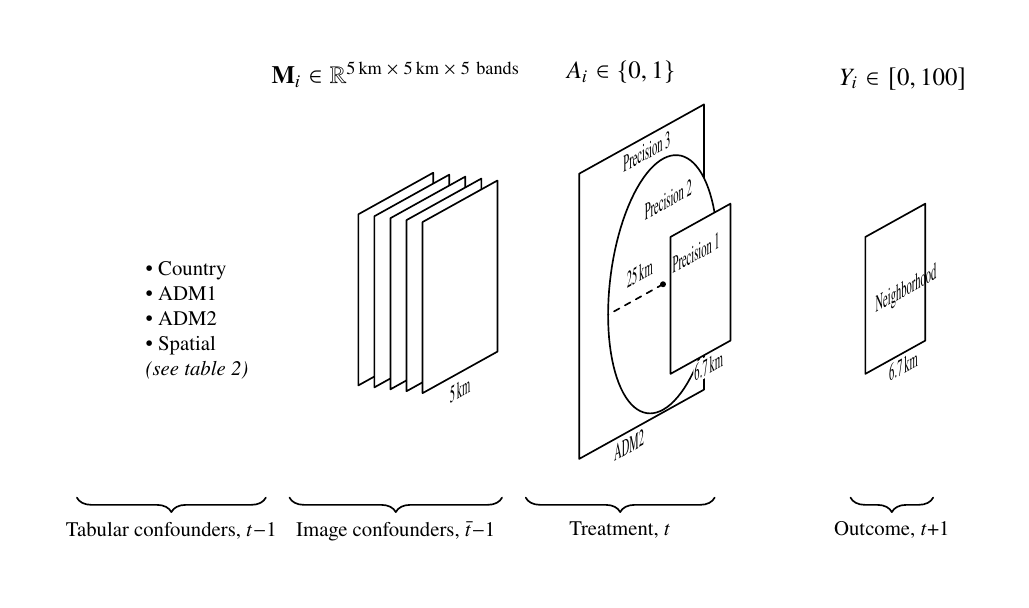}
  \caption{
 Visualizing the shape of the data objects.
  }
  \label{fig:UnitViz}
\end{figure}

Here, a note on the unit of analysis is warranted. As shown in Figure \ref{fig:UnitViz}, the outcome $Y_{it}$ is a raster of 6.7 km side lengths. The outcome represents socioeconomic well-being (i.e., health and living conditions broadly defined) at time $t$, measured via the International Wealth Index derived from satellite imagery as detailed in \citet{Pettersson2023} and further discussed in the data section. Substantively, this square $i$ measures approximately a \textit{neighborhood}---a block in an urban area or a village in a rural area. The outcome is measured one 3-year-period after the treatment has been deployed, for simplicity, denoted $t+1$. The treatment $A_{it}$ denotes the presence or absence of a development project. 

Each neighborhood has geocoordinates $\mathbf{l}_i \in \mathbb{R}^2$ (centroid of the square) and binary treatment indicators $A_{i,k,WB} \in \{0,1\}$ for World Bank projects and $A_{i,k,China} \in \{0,1\}$ for Chinese projects, where $k$ indexes sector-specific interventions (indices for actors and sectors are dropped for brevity where unambiguous). Sector definitions appear in the data section.

A unit is in the treated group if at least one of the following holds for projects in its vicinity: (i) for precision‑1 (exact) locations, the project point lies inside the 6.7 km neighborhood square to capture direct local exposure; (ii) for precision‑2 (``near'') locations, the neighborhood centroid is within 25 km of the reported project point to account for greater spatial uncertainty; or (iii) for precision‑3 locations, where AidData only reports the second administrative division (ADM2), the neighborhood falls inside that ADM2. In other words, we only treat entire ADM2s when project locations are intrinsically coarse (precision 3); for projects with more precise coordinates, we do not treat ADM2s as a whole. Treatment status is determined contemporaneously with the commitment period.\footnote{Investigators who wish to apply alternative treatment-footprint rules (e.g., different buffer radii or point-in-polygon criteria) and assess the sensitivity of estimates can do so using our replication package.} 

Under the potential outcomes framework \citep{Rubin1990}, each unit possesses counterfactuals $Y_{it}(1)$ and $Y_{it}(0)$.  Our analysis targets sector-specific average treatment effects (ATEs) for World Bank and Chinese projects:
\[
\text{ATE} = \mathbb{E}[Y_{it}(1) - Y_{it}(0)].
\]
For a given sector, there will be two ATEs, one for World Bank and another for Chinese projects.

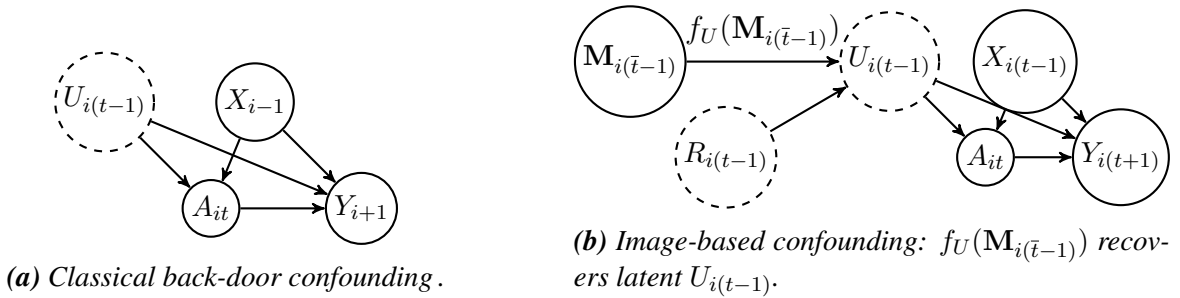
\begin{figure}[ht]
  \centering
  \begin{subfigure}{0.48\textwidth}
    \centering
    \tikzstyle{main node}=[circle,font=\sffamily\small\bfseries,inner sep=1.5pt]
    \tikzstyle{sub node}=[circle,dashed,font=\sffamily\small\bfseries,inner sep=1.5pt]
    \begin{tikzpicture}[->,>=stealth',auto,node distance=2cm,thick]
      \node[sub node] (U) {
      \color{gray} $U_{i(t-1)}$
      };
      \node[main node] (A) [below right of=U] {$A_{it}$};
      \node[main node] (Y) [right of=A] {$Y_{i+1}$};
      \node[main node] (X) [right of=U] {$X_{i-1}$};
      \path
        (U) edge (A)
        (U) edge (Y)
        (A) edge (Y)
        (X) edge (A)
        (X) edge (Y);
    \end{tikzpicture}
    \caption{Classical back‐door confounding\,.}\label{fig:SimpleDag}
  \end{subfigure}%
  \hfill
  \begin{subfigure}{0.48\textwidth}
    \centering
    \tikzstyle{main node}=[circle,font=\sffamily\small\bfseries,inner sep=1.5pt]
    \tikzstyle{sub node}=[circle,dashed,font=\sffamily\small\bfseries,inner sep=1.5pt]
    \begin{tikzpicture}[->,>=stealth',auto,node distance=1.8cm,thick]
      \node[main node] (M)                   {$\mathbf{M}_{i(\overline{t}-1)}$};
      \node[sub node]  (U) [right of=M,xshift=1.65cm]      {
      \color{gray} 
      $U_{i(t-1)}$
      };
      \node[main node] (A) [below right of=U] {$A_{it}$};
      \node[main node] (Y) [right of=A]      {$Y_{i(t+1)}$};
      \node[main node] (X) [right of=U]      {$X_{i(t-1)}$};
      \node[sub node]  (R) [below right of=M] {
      \color{gray} 
      $R_{i(t-1)}$
      };
      \path
      (M) edge node[above] {
      {\scriptsize $f_U(\mathbf M_{i(\overline{t}-1)})$}
      } (U)
        (R) edge (U)
        (U) edge (A)
        (U) edge (Y)
        (A) edge (Y)
        (X) edge (A)
        (X) edge (Y);
    \end{tikzpicture}
    \caption{Image‐based confounding: 
    \(f_U(\mathbf{M}_{i(\overline{t}-1)})\) recovers latent \(U_{i(t-1)}\).}\label{fig:ImageDag}
  \end{subfigure}
  \caption{%
    {\bf (a)} Traditional DAG with unobserved confounder \(U_i\). 
    {\bf (b)} With satellite image \(\mathbf{M}_{i(\overline{t}-1)}\), a function \(f_U(\cdot)\) maps \(\mathbf{M}_{i(\overline{t}-1)}\) into the latent confounder \(U_{i(t-1)}\), blocking the back‐door path when conditioned upon. 
    \(X_{i(t-1)}\) are observed covariates and \(R_{i(t-1)}\) are confounders not present in the image.%
  }\label{fig:IntroDiag}
\end{figure}

Identification commonly proceeds via backdoor adjustment, as depicted by the directed acyclic graph (DAG) in Figure \ref{fig:ImageDag}. Confounding emerges when unobserved factors $U_{i(t-1)}$ influence both assignment $A_{it}$ and outcome $Y_{it}$. In aid allocation, such $U_{i(t-1)}$ often encompasses geospatial attributes---terrain texture, infrastructure layouts, proximity to natural features, and other strategic reasons---that donors consult via maps during site selection. These maps, unavailable to researchers, encode spatiotemporal selection logics that drive why certain neighborhoods receive aid over others. Satellite images $\mathbf{M}_{i(\overline{t}-1)}$, which capture both observable and latent geospatial traits (e.g., inferred historical condition on the ground), serve as proxies for these maps. Conditioning on $\mathbf{M}_{i(\overline{t}-1)}$ thus blocks confounding paths by approximating the assignment mechanism donors employ. The image $\mathbf{M}_{i(\overline{t}-1)}$, as discussed in Data, is measured as a composition of the previous three-year images, prior to the treatment (hence, the bar $\overline{t}$ over the time indicator in Figure \ref{fig:UnitViz}). 

While our image-based adjustment improves identification by recovering elements of $U_{i(t-1)}$, residual confounding $R_{i(t-1)}$---political, cultural, or economic factors beyond imagery or measured tabular covariates---may persist \citep{pearl2009causality}. If $R_{i(t-1)}$ contributes minimally relative to $\mathbf{M}_{i(\overline{t}-1)}$ and observed covariates $\mathbf{X}_{i(t-1)}$, bias remains small.

For intuition, contrast the standard tabular-confounding DAG (Figure \ref{fig:SimpleDag}) with our image-augmented version (Figure \ref{fig:ImageDag}). In the former, unobserved $U_{i(t-1)}$ biases estimates absent adjustment. In the latter, $\mathbf{M}_{i(\overline{t}-1)}$ precedes $U_{i(t-1)}$, as map features shape donor decisions \citep{Ohler2019}. These features are extracted with the help of a computer-vision function $f_U(\mathbf{M}_{i(\overline{t}-1)})$ which estimates what visible and latent structures are relevant for the presence or absence of an aid project. This estimation then enhances backdoor closure when conditioned alongside $\mathbf{X}_{i(t-1)}$.

Adjusting for high-dimensional images $\mathbf{M}_{i(\overline{t}-1)}$ demands specialized methods. We employ a computer vision method to yield $f_U(\mathbf{M}_{i(\overline{t}-1)})$, integrated with $\mathbf{X}_{i(t-1)}$ for propensity estimation and inverse probability weighting. This not only facilitates ATE recovery but elucidates assignment by revealing geospatial drivers of aid placement. This approach is further detailed in Methods. 

Building on this foundation, we will examine variation across funders and sectors in the relative influence of neighborhood-level factors (proxied by imagery) versus structural or institutional factors (proxied by country-level features in $\mathbf{X}_{i(t-1)}$) on assignment decisions. Such analyses promise clearer insights into donor-specific allocation logics, sharpening our understanding of aid's causal pathways.

We will later benchmark image‑based adjustment against simpler specifications (difference‑in‑means, tabular‑only, and ADM2 fixed effects) as well as unit-level fixed effects. 

\subsection{Outcome:  Pettersson et al. (2023)'s Satellite-Derived Wealth Measure}

Our second contribution addresses measurement of socioeconomic well-being (our outcome) by using the IWI data generated by Pettersson et al. (2023). The unit resolution is 6.7 km square and measured in three-year periods (1990-1992, 1993-1995, etc.) over all African human settlement areas, which we call \textit{neighborhoods}. This dataset used three-year intervals to handle clouds and missing images. Pettersson et al. trained a residual neural network machine learning algorithm to associate daytime and nighttime satellite imagery with IWI wealth measures from 138 Demographic and Health Surveys conducted in African countries between 1990 and 2020; Data of 57,195 survey clusters were used for training, which are representative of about 1.2 million households \citep{Pettersson2023}.

The IWI measure, developed by \citet{Smits2015} and building further on the DHS original wealth index \citep{rutstein2015steps}, is a continuous measure from 0 to 100 computed by taking the first principal component from answers to the DHS questionnaire about household access to a TV, refrigerator, phone, bike, car, utensils, and electricity. The measure also includes answers to questions about the quality of water, toilet, and floor facilities in the household, as well as the number of bedrooms \citep{Pettersson2023,Smits2015}. 

\begin{table}[htbp]
\centering
\caption{IWI by year group for  \texttt{dhs\_id = 48923}: With vs. without CV imputations.}\label{tab:DataComp}
\label{tab:iwi_na_side_by_side}
\footnotesize
\begin{minipage}{0.48\linewidth}
\centering
\caption*{(a) Raw data without CV imputations}
\begin{tabular}{llcc}
\toprule
DHS ID & Year group & Treated & IWI \\
\midrule
48923 & 2002:2004 & 1 & ? \\
48923 & 2005:2007 & 0 & 19.6 \\
48923 & 2008:2010 & 0 & ? \\
48923 & 2011:2013 & 1 & ? \\
\bottomrule
\end{tabular}
\end{minipage}
\hfill
\begin{minipage}{0.48\linewidth}
\centering
\caption*{(b) With CV imputations}
\begin{tabular}{llcc}
\toprule
DHS ID & Year group & Treated & IWI \\
\midrule
48923 & 2002:2004 & 1 & 15.1 \\
48923 & 2005:2007 & 0 & 19.6 \\
48923 & 2008:2010 & 0 & 20.3 \\
48923 & 2011:2013 & 1 & 21.0 \\
\bottomrule
\end{tabular}
\end{minipage}
\end{table}

To select neighborhoods, we use the locations of survey clusters from a round of the DHS for each country, as they are a nationally representative random sample of census enumeration areas. They approximate the size of a rural community or urban neighborhood, selected within urban and rural strata \citep{Croft2018}. At the time surveys were conducted, each cluster’s GPS coordinates were collected and then randomly displaced for privacy purposes—by up to 2 km in urban areas and 5 km in rural areas, with an additional 1\% of rural clusters displaced up to 10 km \citep{Burgert2013}. Excluding 11 DHS cluster locations that lacked adequate data, such as those with duplicate latitude and longitude coordinates or no population density estimates, we were left with \GlobalNUnits{} units of analysis. More details and descriptive statistics are provided in Tables DA1 and DA2 in the Data Appendix; see Table \ref{tab:DataComp} for illustration. 

Our baseline outcome definition links each DHS cluster to the corresponding 6.7 km $\times$ 6.7 km IWI tile. This resolution is broadly comparable to the DHS GPS displacement used for confidentiality (up to 2 km in urban areas and up to 5 km in rural areas), but it does not guarantee that the undisplaced cluster location falls within the same tile. To assess sensitivity to this potential spatial misalignment, we also re-estimate our main models using a coarser outcome that averages the four adjacent IWI tiles (an effective 13.4 km $\times$ 13.4 km neighborhood; see Section~\ref{s:raster_robustness}).

\section{Study Setting, Data, and Measurement}

\subsection{Treatments:  World Bank and China Development Projects}

\begin{figure}[htb]
  \centering
  \includegraphics[width=0.75\linewidth]{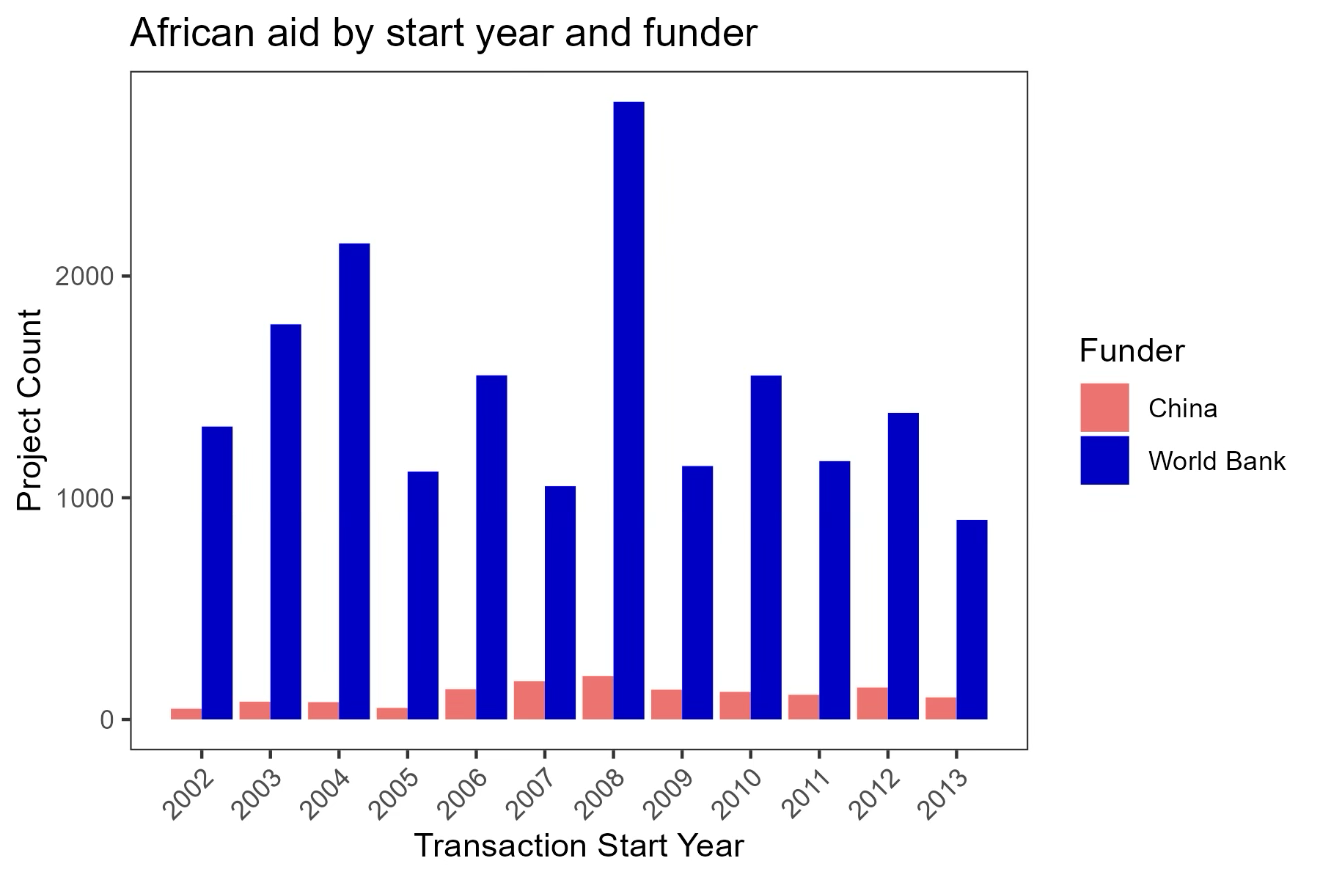}
  \caption{
African aid by start year and funder from 2002-2013.
  }
  \label{fig:AidStartYearFunder}
\end{figure}

\begin{figure}[htb]
\centering
  \includegraphics[width=0.75\linewidth]{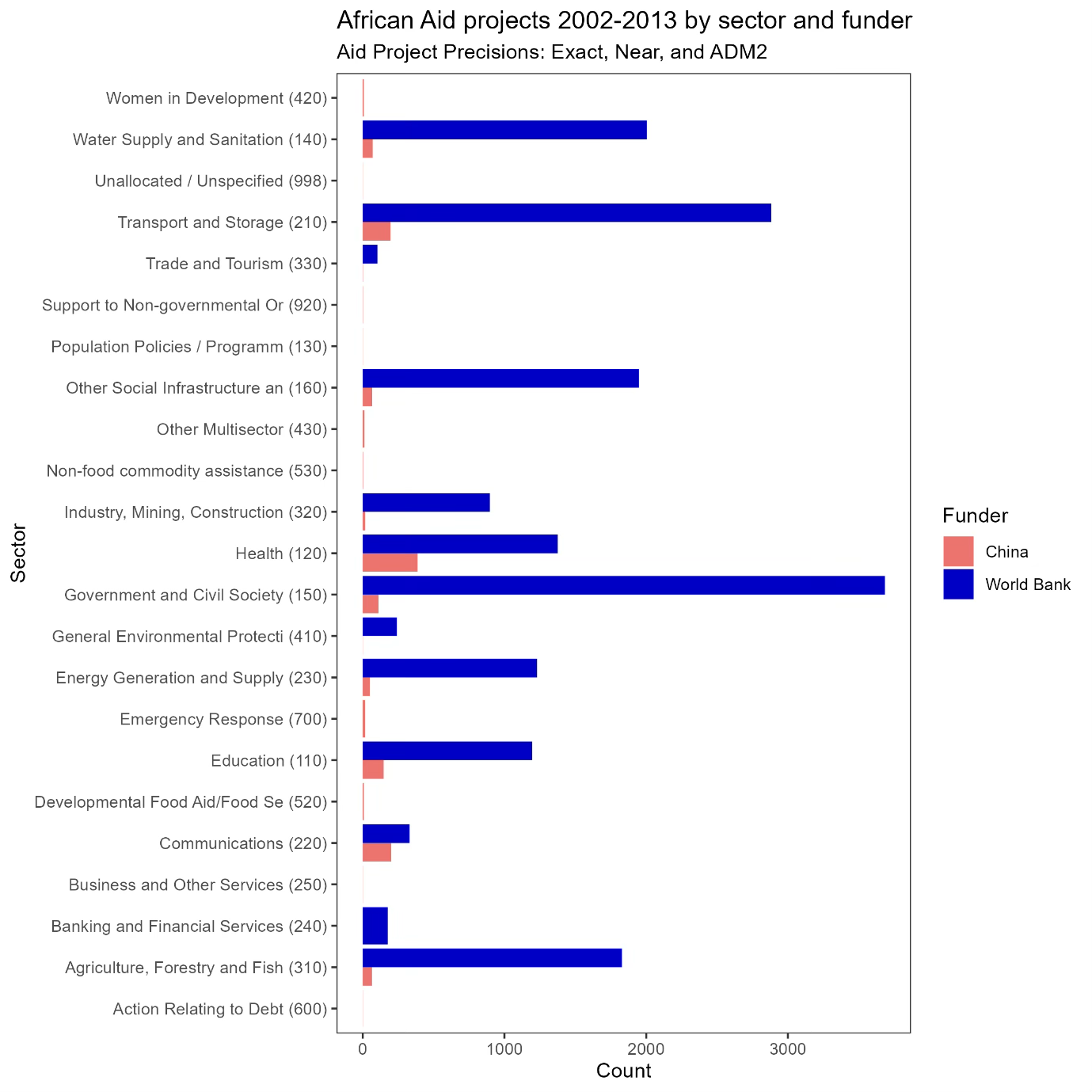}
  \caption{
 African aid projects 2002-2013 by sector and funder.
  }
  \label{fig:SectorFig}
\end{figure}

Having discussed our outcome data in the previous section, we now discuss the treatment. Data on both World Bank and China aid projects were sourced from \citep{Dreher2021b} and curated by AidData.  Table \ref{tab:nProjects} shows descriptive statistics, Figure \ref{fig:AidStartYearFunder} illustrates annual project counts, and Figure \ref{fig:SectorFig} depicts the sector distribution for the two funders over the study period.\footnote{Aid projects spanning multiple sectors are treated by allocating the full project count to each relevant sector or subsector, without weighting or fractional division based on the number of sectors involved.} The general takeaway is that China funds more projects overall (Table~\ref{tab:nProjects}), but the World Bank’s portfolio spans many more distinct project types and locations.

\paragraph{World Bank Projects.}

The World Bank makes its activities publicly available; however, to obtain the latitude and longitude of project locations along with a location precision code, we use AidData’s version 1.4.2 dataset \citep{AidData2017}. We use only projects started between 2002 and 2013, as image and confounder data are missing for earlier years, and project data are missing for 2015-2016 in the 3-year group, including 2014 projects. For projects labeled with multiple sectors, we create an observation for each parent sector (e.g., ``110-Education'').  We limit the data to projects in African countries hosting DHS surveys.

\begin{table}
\footnotesize\centering
\begin{tabular}{lrr}
\toprule
 & China & World Bank \\
\midrule
Countries hosting projects count & 50 & 44 \\
Sectors funded & 22 & 13 \\
Aid project count & 722 & 513 \\
Aid project location count & 1{,}373 & 7{,}115 \\
Exact locations available (precision 1) & 987 & 4{,}149 \\
Near ($<25$km) locations available (precision 2) & 169 & 258 \\
ADM2 locations available (precision 3) & 217 & 2{,}708 \\
Portion lacking end date & 68\% & 15\% \\
Portion lacking full funding information & 100\% & 28\% \\
Portion with concurrent, co-located, multi-sector projects & 3\% & 2\% \\
\bottomrule
\end{tabular}
\caption{
Comparison of China and World Bank aid in Africa from 2002 to 2013.
}\label{tab:nProjects}
\end{table}

\paragraph{Chinese Aid Projects.} 
China does not publicly share data about its aid projects, but AidData used the Tracking Underreported Financial Flows (TUFF) methodology to gather this data from four types of open sources:  media reports, official statements from Chinese officials and organizations, aid and debt information systems in recipient countries, and research from scholars and non-governmental organizations \citep{AidDataResearchandEvaluationUnit2017,Strange2017}. We use version 1.1.1 of their Global Chinese Development Finance dataset.

We exclude pre-2002 and post-2013 projects due to missing satellite and covariate data for earlier periods, as well as missing project data in 2015-2016 for the 3-year group, which includes projects from 2014. We limit projects to African locations using a process described in the Data Appendix.  We include projects labeled as ``year uncertain'' by AidData because they used the year of the first source mentioning the project. We include only projects in completed or implementation stages at the time of dataset publication, whose funding types and objectives met the OECD definition for Official Development Assistance (ODA), making them comparable to World Bank aid.  Just like with World Bank aid, we include only locally targeted aid (location precisions 1-3). 

\paragraph{Defining Treatment/Control and Creating Panel Data.}

As discussed previously and shown in Figure \ref{fig:UnitViz}, to define treatment and control groups and create panel data for each funder and sector, we define a neighborhood unit as part of the treatment group if any of the following three conditions is met: a precision 1 (exact) project lies within the \SpatialRes{} km neighborhood boundary, the neighborhood lies within a 25 km buffer around a precision 2 (near) project boundary, or the neighborhood lies within an all-ADM2 (precision 3) project area.\footnote{
Note that the ADM2‑based rule applies only to precision‑3 locations; when AidData reports precision‑1 or precision‑2 coordinates, we rely solely on the \SpatialRes{} km and 25 km square rules and do not treat entire ADM2s.
} If none of these conditions are fulfilled, the unit is assigned to the control group. To avoid confounding at the country selection level, we exclude neighborhoods from the control group in countries where the funder did not conduct any projects in the sector of interest during the study period.

Focusing on these higher-resolution locality resolutions, rather than country- or ADM1-level resolutions, is better aligned with how local aid projects are deployed and their efficiency compared to country assignments \citep{Dreher2015a,Winters2010}. Additionally, focusing on local aid enables us to leverage the proxy maps that World Bank and Chinese officials may have been using in allocating aid---unobserved by researchers---and proxied by satellite imagery. 

\begin{figure}
    \centering
    \includegraphics[width=1\linewidth]{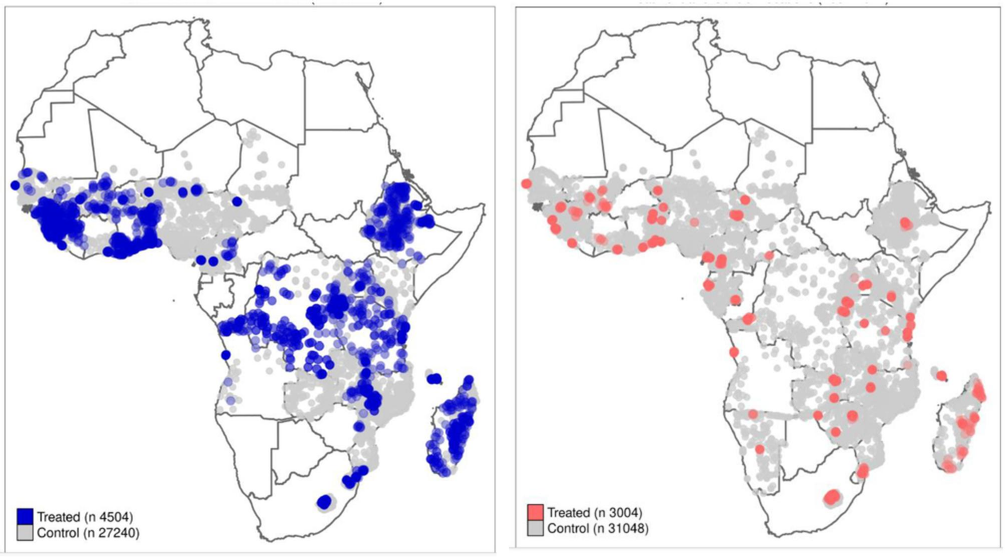}
    \caption{
    Geographical distribution of World Bank and Chinese projects. Note that some aid projects cannot be distinguished visually at the continental scale.
    }
    \label{fig:placeholder}
\end{figure}

After identifying the treatment status for all neighborhoods each period, we create panel data of control and treated groups with three-year observations by funder and sector. Figure \ref{fig:placeholder} maps treated and control neighborhoods for World Bank and Chinese projects.  Maps for all funders by sector are available in the Data Appendix.

We exclude funder-sector combinations with fewer than 80 treated locations, as they lack sufficient data for training and test splits.  For the remaining sector/funder combinations, we check for and remove variables that lack variation across the treatment and control groups, as they add little statistical value.  

Lastly, we estimate the ATE of each sector and funder’s aid one 3-year lag after the 3-year project commitment period. This lag is important because subnational Chinese aid studies found the strongest effects during that period: Dreher et al. (2021) in year 3 and \citep{Xu2020} two years after commitment. 

\subsection{Tabular-Covariates Adjustment Set}

We provide a large set of tabular confounders that the literature suggests influence both aid placement and effectiveness, variables for year fixed effects to account for common shocks, and ADM2 variables for fixed effects to account for numerous time-invariant factors at that level, including regional history, ethnic composition, and culture.  There may be additional unobserved confounders that threaten inference. To avoid introducing post-treatment bias, covariates are measured pre-treatment where possible, making it more likely that they function as good controls (see \citet{cinelli2024crash} for a detailed discussion). Tabular variable measurement is summarized in Table \ref{tab:Descriptives} and the section below. The Data Appendix contains more details.

\singlespacing
{ \footnotesize
\begin{longtable}{p{5.5cm} p{3cm} p{6.5cm}}
  \caption{\it 
  Tabular covariates.}\label{tab:Descriptives}
  \label{tab:covariates} \\

  \toprule
  \textbf{Variable} & \textbf{Resolution} & \textbf{Time} \\
  \midrule
\endfirsthead

\caption*{\it Covariate descriptions (continued).} \\ 
  \toprule
  \textbf{Variable} & \textbf{Resolution} & \textbf{Time} \\
  \midrule
\endhead

  \midrule
  \multicolumn{3}{r}{\footnotesize\textit{Continued on next page}} \\
\endfoot

  \bottomrule
\endlastfoot

    \rowcolor{gray!25}\multicolumn{3}{l}{\textbf{Time-varying country-level covariates}} \\

   $\cdot$ UN Security Council Temporary Member aligned with U.S. indicator
      & Country
      & 1 if was a temporary UNSC member and always voted with the U.S.\ in any of the 3 years prior to the project commitment period, 0 otherwise. \\

 $\cdot$   UN Security Council Temporary Member non-aligned with U.S. indicator
      & Country
      & 1 if was a temporary UNSC member and voted against the U.S.\ at least once in any of the 3 years prior to the project commitment period, 0 otherwise. \\

  $\cdot$  Gini inequality measure
      & Country
      & 3‑year average, one period prior to project commitment \\

  $\cdot$  Control of corruption score from annual Worldwide Governance Indicators
      & Country
      & 3‑year average, one period prior to project commitment \\

   $\cdot$ Government effectiveness score from annual Worldwide Governance Indicators
      & Country
      & 3‑year average, one period prior to project commitment \\

   $\cdot$ Political stability and absence of violence/terrorism score from annual Worldwide Governance Indicators
      & Country
      & 3‑year average, one period prior to project commitment \\

$\cdot$    Regulatory quality score from annual Worldwide Governance Indicators
      & Country
      & 3‑year average, one period prior to project commitment \\

    $\cdot$ Rule of law score from annual Worldwide Governance Indicators
      & Country
      & 3‑year average, one period prior to project commitment \\

    $\cdot$ Voice and accountability score from annual Worldwide Governance Indicators
      & Country
      & 3‑year average, one period prior to project commitment \\

    \midrule

    \rowcolor{gray!25}\multicolumn{3}{l}{\textbf{Time-varying covariates of source spacecraft for composite daytime images}} \\

    $\cdot$ Landsat satellite spacecraft category
      & 3‑year period
      & 1 for periods 2011 and later, when images are a composite of Landsat 5, 7, and 8. Earlier periods are assigned 0, when images are composites of Landsat 5 and 7. \\

    \midrule

    \rowcolor{gray!25}\multicolumn{3}{l}{\textbf{Fixed Effects Covariates}} \\

    $\cdot$ Year Group
      & 3‑year period
      & First year of group included in each 3‑year observation \\

    $\cdot$ Neighborhood’s second administrative division (ADM2)
      & ADM2
      & Time invariant \\

    \midrule

    \rowcolor{gray!25}\multicolumn{3}{l}{\textbf{Covariates for each neighborhood location}} \\

    $\cdot$ Nightlights, per capita, logged
      & \SpatialRes{} km 
      & 3‑year average, one period prior to project commitment \\

    $\cdot$ Population density, average of 1 km squares within neighborhood, logged
      & \SpatialRes{} km 
      & 3‑year average, one period prior to project commitment \\

    $\cdot$ Distance to nearest known deposits of Gold, Gems, Diamonds, and Oil, kilometers, logged
      & Point-to-point distance
      & 3‑year average, one period prior to project commitment \\

    $\cdot$ Travel minutes to city of over 50,000 population in the year 2000, average of $\sim$1 km squares in neighborhood, logged
      & \SpatialRes{} km 
      & Time invariant \\

    \midrule

    \rowcolor{gray!25}\multicolumn{3}{l}{\textbf{Time-varying covariates, inherited from neighborhoods’ administrative 1, 2, or 3 area}} \\

    $\cdot$ Concurrent aid project count (both funders / any sector) in adjacent ADM2s, logged
      & ADM2
      & During project commitment period \\

    $\cdot$ Current funder’s concurrent project count in other sectors, logged
      & ADM2
      & During project commitment period \\

    $\cdot$ Other funder’s concurrent project count in all sectors, logged
      & ADM2
      & During project commitment period \\

    $\cdot$ Chinese loan-based project count
      & ADM1 and ADM2
      & During project commitment period \\

    $\cdot$ Birthplace of country’s executive leader in power indicator
      & ADM1
      & 1 if was leader birthplace in any of the 3 years in pre‑project commitment period, 0 otherwise. \\

    $\cdot$ Conflict death count, logged
      & ADM1
      & 3‑year sum, one period prior to project commitment \\

    $\cdot$ Natural disaster count, logged
      & ADM1, ADM2, and ADM3
      & 3‑year sum, one period prior to project commitment \\

    \midrule

    \rowcolor{gray!25}\multicolumn{3}{l}{\textbf{Time-varying covariates, inherited from country}} \\

$\cdot$ Election year for national executive indicator
      & Country
      & 1 if any of the 3 years in pre‑project commitment period were an election year, 0 otherwise. \\

\end{longtable}
}
\defaultspacing

\paragraph{Per-capita Nightlights.}

Because previous literature suggests both World Bank and China place projects in the non-poorest sub-national locations \citep{Briggs2021,Dreher2019} and that projects are most successful in areas with brighter night lights (conditional on population) and above-average education \citep{Briggs2021,Greer2021a}, we calculate a per-capita nightlight variable to represent a pre-project neighborhood wealth confounder. We use nightlight data from the Defense Meteorological Satellite Program Operational Linescan System (DMSP) and population data from Columbia University’s WorldPop Global Project \citep{WorldPopPopulationDensitynd}.  For each three-year period, we average the annual per capita nightlights and then add one, taking the log because the data are highly right-skewed.   

Because nightlights are persistent and also enter the Pettersson et al.\ (2023) outcome-construction pipeline, we also conduct robustness checks that drop lagged per-capita nightlights from the adjustment set. Results are reported in Section~\ref{s:NTLrobustness}.

\paragraph{Population Density.}

Since empirical research suggests both China and the World Bank allocate aid to more populous subnational regions \citep{Dreher2019,Dreher2022,Ohler2019} and population density can facilitate economic opportunity and market access \citep{WorldBank2008}, we create an annual average population density variable for the square areas around each unit of observation neighborhood from the WorldPop Population Density dataset \citep{WorldPopPopulationDensitynd}. Since the average population density is highly skewed, we add one and take the log after averaging the value over 3-year periods. 

\paragraph{Distance to Gold, Gems, Diamonds, and Petroleum.}

Past studies have found a relationship between project locations and natural resources \citep{Broich2017,Guillon2020}, and although one study did not \citep{Dreher2015}, we include as covariates each DHS cluster location’s distance to known deposits of four natural resources:  gold, gems, diamonds, and petroleum, which are updated over the study period as new deposits were discovered. 

We calculate each neighborhood’s distance in kilometers from resource deposits known in each period with the following datasets: GOLDDATA 1.2 (Balestri, 2012), GEMSTONE (Flöter et al., 2005), DIADATA \citep{Gilmore2005}, and PETRODATA \citep{Lujala2007}. Because the data is highly skewed, we add one and take the log.

\paragraph{Travel Minutes to City of Over 50,000 Population in the Year 2000.}

Because empirical studies of both World Bank and China aid in Africa suggest funders avoid areas with high travel times to a city \citep{Dreher2022}, and access to an urban area can provide economic opportunities that increase aid effectiveness \citep{WorldBank2008}, we calculate each neighborhood’s travel minutes to a city. To operationalize this, we use data on the estimated travel time to the nearest city of 50,000 or more inhabitants in the year 2000 \citep{Nelson2008} to calculate the travel time in minutes for each point within the \SpatialRes{} km square area surrounding each unit of observation. We take the average for the square area, add one, and take the log.    

\paragraph{Birthplace of Leaders in Power.}

Because previous research suggests China gives more aid to provinces in years when they are the birthplace of the country’s leaders than in other years \citep{Dreher2019}, we use tabular, geocoded data on the birth region of political leaders and the years they were in power available from the Political Leaders' Affiliation Database (PLAD) \citep{bomprezzi2024wedded}. We set the variable to 1 if the national executive leader in power during the annual or 3-year period was born in the neighborhood’s ADM1 and to 0 otherwise.  

\paragraph{Nearby Chinese Loan-based Projects.}

Because China co-locates aid and loan projects to achieve agglomeration effects to increase the economic impact of their aid, we create a count of annual Chinese loan projects in the same ADM1 or ADM2 as the DHS cluster location, from AidData data about Chinese Other Official Flows (OOF) available in their version 1.1.1 dataset \citep{Strange2017}.   Because the data is skewed, we add one and take the log.

\paragraph{Treated Other Funder Count.}

Because simultaneous aid activities from China influence economic growth in locations of active World Bank projects, and vice versa, we create a variable to count the other funder’s simultaneous projects in all sectors within the same ADM2 as the unit of observation, the neighborhood.  While it would have been ideal to have a measure of the dollar amounts invested by the other funder, this data is not available for all Chinese projects, so we use a count instead.  Because the data is skewed, we add one and take the log.  

\paragraph{Other Sector Project Count.}

Because aid projects the funder does in other sectors can influence the effectiveness of the sector under study, such as transportation projects enabling easier access to Health and Education projects, and because China co-locates these projects to achieve agglomeration effects to improve economic growth, we create a variable to count the funder’s projects in other sectors in the same ADM2 as the unit of observation neighborhood.  For multi-sector World Bank projects, this variable is incremented once for each sector except the one being studied.  While it would have been ideal to measure dollar amounts invested, this data is not available for all Chinese projects, so we use a count instead. Because the data is skewed, we add one and take the log. 

\paragraph{Adjacent ADM2 Project Count.}

Because the effects of aid projects in adjacent ADM2 regions may spill over to the ADM2 hosting the neighborhood of analysis \citep{Bitzer2018,Demir2023,Xu2020}, we create a variable to count concurrent World Bank and Chinese aid projects in all sectors in all ADM2s adjacent to the one hosting the neighborhood of analysis, including those across country borders.  Because the data is skewed, we add one and take the log.  

\paragraph{Conflict Deaths.}

Studies find China reduces aid in response to conflict \citep{Sardoschau2019} and find mixed results on the World Bank’s response to conflict \citep{BenYishay2022,Ohler2014}.  Because conflicts can increase the need for aid and the difficulty of projects, we include battle-related deaths from the Uppsala Conflict Data Program’s Georeferenced Event Dataset \citep{Croicu2015,Gehring2022}.  The variable is a 3-year sum of conflict deaths in the neighborhood’s ADM1; because it is skewed, we add one and take the log.

\paragraph{Natural Disasters.}

Because natural disasters impact both the need for aid and a region’s aid absorption ability \citep{Drabo2021}, we create a natural disaster count for each neighborhood and period that includes ``floods, storms, earthquakes, volcanic activity, extreme temperatures, landslides, droughts, and (dry) mass movements'' \citep[p.~2]{Rosvold2021}.  The disaster data is from the International Disaster Database (EM-DAT) \citep{Delforge2023}, and geocoding is from the Center for International Earth Science Information Network’s GDIS dataset \citep{Rosvold2021}. Because the data is highly right-skewed, we add one and take the log. 

\paragraph{Election Year Indicator.}

Since empirical research indicates more political capture of Chinese aid happens during election years \citep{Dolan2020,Dreher2019,Dreher2022}, we create an indicator of whether national executive elections were held each period, based on the Database of Political Institution's \textit{Years left in current term} variable, which is set to 0 for election years of national executive leaders.   
\paragraph{UN Security Council Temporary Membership and Voting}

Since empirical research suggests both World Bank and Chinese aid allocation are influenced by recipient countries’ temporary membership in the UN Security Council and their votes there \citep{Berlin2023,Dreher2018a,Guillon2020}, we create two variables to represent UN Security Council votes and temporary membership.  One variable indicates complete alignment with the U.S. position, and the other indicates at least one deviating vote during the period; non-members are assigned a value of 0. This all-or-nothing operationalization approach follows \citet{Berlin2023} and \citet{Dreher2018a}, who suggest the U.S. may punish a deviating vote by withholding World Bank aid.  Neighborhoods inherit variables from their country.  The UNSC's temporary membership data were sourced from \citep{Dreher2009}, and the UNSC's voting data were obtained from \citet{Dreher2022a}.

\paragraph{Country-level Governance Indicators.}

A large body of literature argues that aid allocation and effectiveness depend on the quality of recipient country institutions (see \citet{Asatullaeva2021} for a review).  To capture multiple dimensions of recipient country governance and institutions, we use six country-level, annual variables from the World Bank’s Worldwide Governance Indicators \citep{Kaufmann2023}:  Control of Corruption, Government Effectiveness, Political Stability and Absence of Violence/Terrorism, Regulatory Quality, Rule of Law, and Voice and Accountability. Each neighborhood inherits the average score of its country each period, which is roughly normally distributed between -2.5 and 2.5.

\paragraph{Country-level Inequality.}

Since previous research suggests that less income inequality is associated with better aid effectiveness \citep{Skardand}, we include a measure of country-level inequality, as country-level inequality, compared to within ADM2 level inequality, captures the broader institutional and policy environment that influences both aid allocation and the outcomes of development interventions \citep{chong2009can}. Each three-year period, each neighborhood inherits the average of the three national annual Gini inequality scores from the World Bank’s World Development Indicators dataset \citep{GiniCoefficient2023}, which takes on values between 0 and 1.

\subsection{Daytime Satellite Image Confounder Details}

To proxy remaining confounders, we use daytime satellite images of approximately 5 km square centered over each neighborhood’s 6.7 km square center. The images came from the U.S. Geological Survey Landsat 5, 7, and 8 satellite series, but since Landsat 5 data was not saved over large portions of equatorial Africa due to lack of commercial data buyers when it was generated \citep{NASAnd}, we limit our study years to exclude projects started prior to 2002, where only Landsat 5 images are available. 

Due to clouding and since no single Landsat series covers all study years (Landsat 7 images taken on and after May 31, 2003, also have diagonal lines across them due to a Scan Line Corrector (SLC) hardware failure \citep{Landsat7LandsatSciencend}), we create composite images from the median pixel of all available satellites each period. This per-pixel median mosaic across sensors and dates mitigates the Landsat-7 SLC-off striping by infilling the missing scan lines with valid pixels from overlapping acquisitions. For each cluster, we build fixed, non-overlapping three-year composites---1999–2001, 2002–2004, 
2005–2007, 
2008–2010, 
and 2011–2013---taking the per-pixel median of all available, cloud-masked Landsat 5/7/8 scenes in the window; these windows are identical across units and do not depend on treatment timing.   We use visible light bands (red, green, blue) plus near-infrared and short-wave infrared spectral bands to capture information about built-up areas.  More detail is available in the Data Appendix.

Causal inference methods require overlap between the characteristics of treated and control units.  While adding many covariates helps control confoundedness, including too many covariates can reduce the overlap needed to make causal claims. This is especially true for high-dimensional data \citep{DAmour2021}, such as satellite images, which are unique across different locations. Our approach addresses this by creating a lower-dimensional representation of satellite images that captures the key features and characteristics of neighborhoods found across various geographic locations.   

\section{Estimation Method: Image-based Confounding Analysis }\label{s:DataMethods}

Part of our analysis relies on a computer‑vision‑based causal-adjustment method for processing satellite images that we developed elsewhere \cite{Jerzak2023,Jerzak2023a}. We use Vision Transformers, a computer vision machine learning algorithm, for our primary results, which rely on \ImModel{}, because this model is generally considered state-of-the-art (but also reports results for tabular-only, and later, unit fixed effects, analysis).

Relative to this earlier work, our contribution in this paper is to deploy the EO‑ML adjustment framework in an applied setting—a multi‑funder, multi‑sector evaluation—and to add diagnostic layers. In particular, we (i) estimate a set of nested specifications (diff‑in‑means, X+FE, M, M+X+FE) to quantify how much imagery changes estimated ATEs, and (ii) compute assignment‑mechanism diagnostics (out‑of‑sample AUCs and tabular‑covariate salience measures) that characterize how donors condition on visible and tabular features. We treat these steps as methodological validation and refinement rather than as a new estimator.

\begin{figure}[htb]
  \centering
\includegraphics[width=1.1\linewidth]{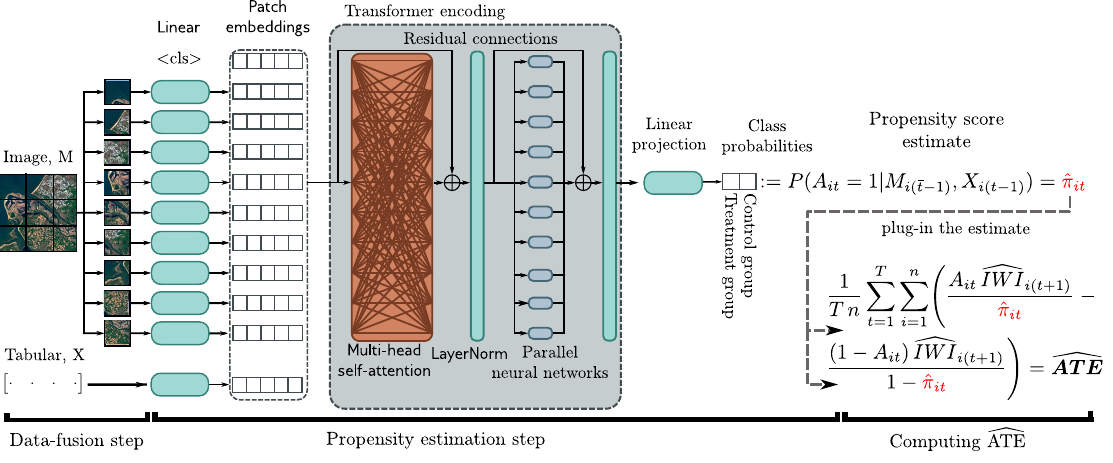}
  \caption{
A stylized visualization of the flow of data fusion, transformer architecture for propensity estimation, and the outcome model. The transformer-block visualization is inspired by \citeauthor{prince2023understanding} 2023. The plug-in estimate, $\widehat{\text{ATE}}$, uses out-of-fold indices via cross-fitting, breaking reuse of the same data for training and estimation, and limiting regularization bias. 
}
\label{fig:TransViz}
\end{figure}

Figure \ref{fig:TransViz} depicts the key intuition behind our causal-\ImModel{} method. The modeling consists of three steps: fusing image and tabular data, processing these data via transformer encoding to estimate the propensity scores, and, finally, plugging those estimates, via cross-fitting, into an inverse probability weighted mean difference estimator of the ATE. The primary objective of the propensity estimation stage is to process images and tabular data to refine our estimation of the propensity weights, which represent the probability of selection into development programs. Our hypothesis is that these weights will better balance our units than tabular weights alone in calculating the mean difference between treated and control units. 

\paragraph{Data-Fusion Step.} After assembling the data, we prepare them for modeling. Besides the usual data curation, this preparation entails assigning all the tabular data in the set $\mathbf{X}_{i(t-1)}$ and the images in the set $\mathbf{M}_{i(\overline{t}-1)}$. The approach here divides each image $i$ into smaller patches for the model to digest efficiently; we use a 16 $\times$ 16 patch size. Then we flatten the patches, preparing them for injection into the model. The tabular data has a different dimensionality, and we must therefore project it to the embedding dimensionality. To this, we also append a summary representation vector called \textit{classification token} (typically denoted CLS). The data streams are now ready for fusion by concatenating the patch, CLS, and (when used) tabular representations.

To summarize, while the tabular covariates contain information about a variety of political, economic, and geographical factors not visible by remote observation, the satellite images contain visible factors relevant for confounding between aid and the wealth outcome. By fusing both sources, the model will learn how to combine the importance of all these factors jointly.  

\paragraph{Propensity-Estimation Step.}

The goals of the propensity-estimation step are to predict treatment status using the images and the supplied tabular data. This step unfolds as follows. It scales and zero-centers the data, and then appends the image pixel values to the tabular confounding variables, creating high-dimensional embeddings where the number of covariates is significantly greater than the number of observations. It regularizes the data to avoid overfitting. Then, these embeddings pass through our computer vision method to learn informative summaries of the raw image and tabular variables, enabling comparison between key features of treated and control neighborhoods. 

Our primary EO-ML method is the  \ImModel{}, and thus passes these embeddings through a transformer encoder. There are many insightful introductions to transformer models \cite{prince2023understanding}, including the original article \citep{vaswani2017attention}. However, here we provide a high-level intuition for this encoder. 

\ImModel{}s capture global context by modeling long-range dependencies across image patches via self-attention, which allows them to effectively integrate spatially distributed features in satellite images. Readers can think of self-attention as a statistical association model that captures relationships between each image patch with all others in an embedding space. In our Figure \ref{fig:TransViz}, this is visualized through the Multi-head self-attention block. It is \textit{multi-head} because the self-attention mechanism is conducted multiple times independently, ensuring that some heads capture information that other heads might have missed. This information is then passed through a normalization block and subsequently through parallel neural-network layers, allowing for further learning on how the image and tabular data are related to each other and how well they estimate treatment versus control class probabilities. In this strategy, we employ an 8-layer model with dropout and drop path (both at 10\%; \citep{huang2016deep}) and balanced training to improve convergence in the context of imbalanced treatment/control classes \citep{francazi2023theoretical}. 

The end-to-end trained \ImModel{} approach uses a random sample of 90\% of the ADM2 areas to train the model, employing stochastic gradient descent to maximize the model’s log likelihood, updating model weights using backpropagation. We employ a cross-fitting approach, using out-of-fold predicted treatment/control probabilities  \cite{chernozhukov2018double} to improve inference and generate uncertainty estimates.

\paragraph{Computing $\widehat{\textrm{ATE}}$.}

Once the transformer finishes learning the class probabilities, we plug them into an Inverse Probability weighting method with the H\'{a}jek adjustment, normalizing the observation weights to sum to one. As shown in the following formulae, the methods used here rely on estimating the ATE by taking a mean difference in the expected value of the post-aid International Wealth Index estimates between neighborhoods that did and did not receive aid, adjusting for pre-project satellite imagery over treatment and control locations, along with tabular data about each location’s treatment status and confounding variables. Formally, this estimation process contains two stages, as follows:  
\begin{equation}
\label{eq:propensity}
\hat{\pi}_{it}
\;=\;
\widehat{P}\bigl(A_{it}=1 \mid 
\mathbf{M}_{i(\overline{t}-1)},
\,\mathbf{X}_{i(t-1)}
\bigr)
\;=\;
f\bigl(\mathbf{M}_{i(\overline{t}-1)},\,\mathbf{X}_{i(t-1)}\bigr)
\end{equation}
\begin{equation}
\label{eq:ate}
\widehat{\textrm{ATE}}
\;=\;
\frac{1}{T\,n}
\sum_{t=1}^{T}
\sum_{i=1}^{n}
\Biggl(
\frac{A_{it}\,\widehat{\textrm{IWI}}_{i(t+1)}}{\hat{\pi}_{it}}
\;-\;
\frac{(1 - A_{it})\,\widehat{\textrm{IWI}}_{i(t+1)}}{1 - \hat{\pi}_{it}}
\Biggr)
\end{equation}
where
\begin{itemize}
  \item[]\hspace{-1.9cm}
  $\displaystyle 
    \hat{\pi}_{it} =\widehat{P}\bigl(A_{it} = 1 \mid 
    \mathbf{M}_{i(\overline{t}-1)},\,\mathbf{X}_{i(t-1)}\bigr)$
    is the estimated treatment probability of neighborhood $i$ at project commitment time.
      \item[$f(\cdot)$] is the image model for the predicted treatment probabilities.
  \item[$\mathbf{M}_{i(\overline{t}-1)}$] is a 5 km square daytime satellite image centered around neighborhood $i$’s location, one lag prior to project commitment. It is a cloudless composite image spanning a 3-year period; for example, project commitments between 2002 and 2004 are based on images from 1999 to 2001.
  \item[$\mathbf{X}_{i(t-1)}$] are tabular confounders in neighborhood $i$, one lag prior to project commitment, including ADM2 and year fixed effects. These are summarized in Table \ref{tab:Descriptives} and in the Tabular Covariates section below.
  \item[$\widehat{\mathrm{ATE}}$] is the estimated average treatment effect.
  \item[$n$] is the total number of unique neighborhoods for the specific funder and sector.
  \item[$T$] is the total number of time periods.
  \item[$A_{it}$] is the observed treatment status for neighborhood $i$ in period $t$ for the specific funder and sector.
  \item[$\widehat{\mathrm{IWI}}_{i(t+1)}$] is the estimated International Wealth Index in community $i$ (ranges from 0–100), measured in a 3-year composite estimate one lag after aid project commitment. For project commitments between 2002 and 2004, for example, IWI is measured between 2005 and 2007.
\end{itemize}
(Note that H\'{a}jek adjustment modifies this baseline estimator slightly by normalizing the weights for improved efficiency \citep{freedman2008weighting}.

\section{Results}\label{s:Results}

First, we discuss the overlap of treatment propensity scores between treated and control neighborhoods for the two funders, then we present the estimated ATEs by sector. Next, we present model quality measures and a summary of each tabular covariate’s salience to aid allocation decisions across funders and sectors. 

\subsection{Average Treatment Effect by Funder and Sector}

\begin{figure}[htb]
  \centering
  \includegraphics[width=0.99\linewidth]{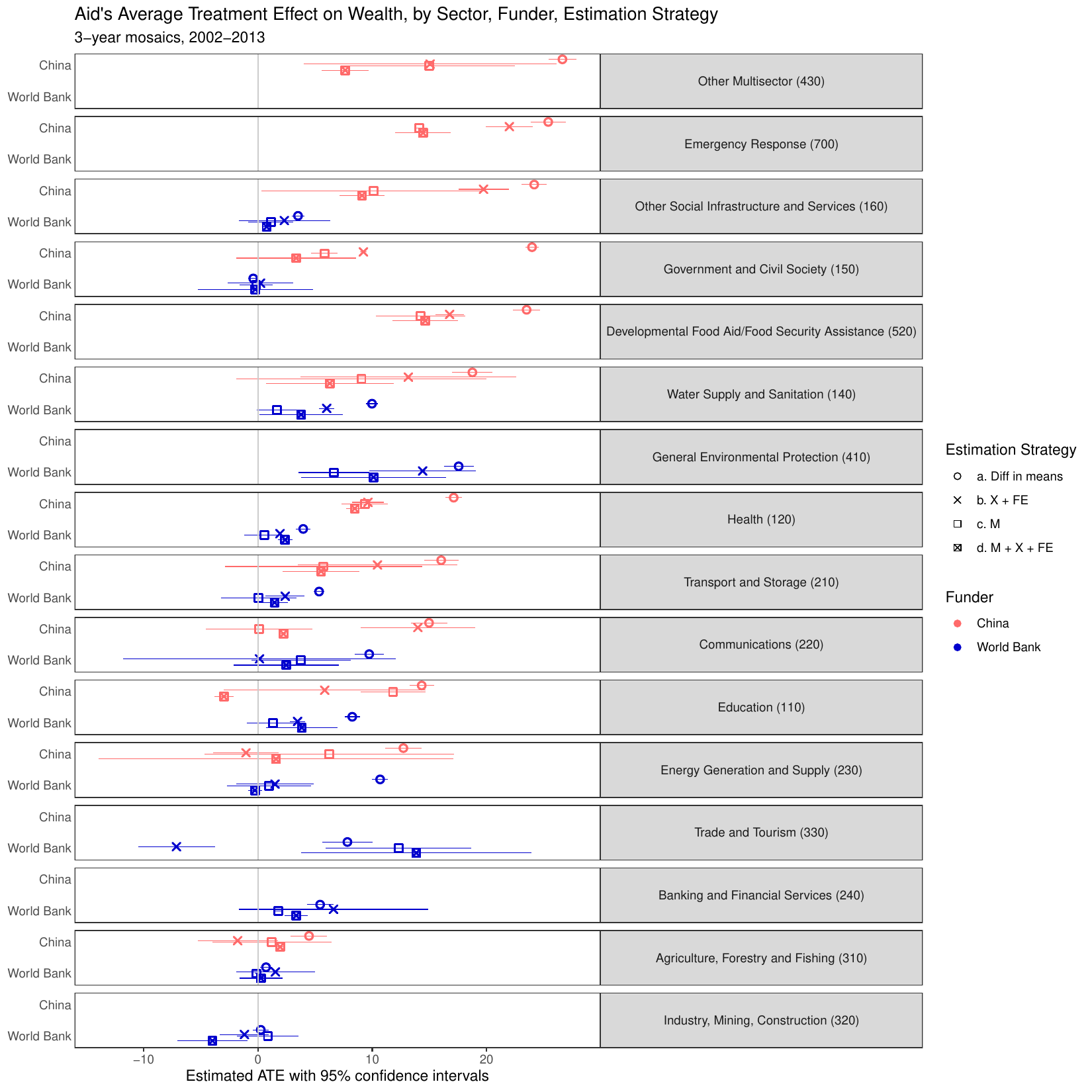}
  \caption{
    ATEs by sector, funder, and modeling strategy. ``FE" refers to ADM2 fixed effects; specifications: (a) Diff‑in‑means, (b) X+FE, (c) M, (d) M+X+FE.
  }
  \label{fig:ATEs}
\end{figure}

Using our \ImModel{}, we estimate ATEs for each funder and report four specifications that correspond to Figure~\ref{fig:ATEs}: 
(a) a difference-in-means baseline, 
(b) tabular covariates with ADM2 fixed effects (X+FE), 
(c) imagery only (M), and 
(d) imagery plus tabular covariates and ADM2 fixed effects (M+X+FE). This subsection also provides our main methodological benchmark: by comparing diff‑in‑means, tabular‑only (X+FE), imagery‑only (M), and imagery‑plus‑tabular (M+X+FE) estimators, we assess how much image‑based confounding control changes estimated ATEs relative to standard approaches. 

Because Figure \ref{fig:ATEs} juxtaposes several estimation regimes, a natural question is which one should be treated as the primary causal estimate. Our preferred adjustment set is the fully image-augmented specification (d), which conditions on pre-treatment satellite imagery—intended to proxy the map-based information donors use when selecting project sites—while also incorporating the standard tabular covariates and ADM2 fixed effects that absorb time-invariant local context.

Across many sectors, adding imagery (c and d) shifts ATEs downward relative to tabular-only or unadjusted models, indicating that satellite images capture additional confounding information that standard covariates may miss. In general, Chinese projects display larger effects than World Bank projects and sectoral heterogeneity. Notably, across sectors, the imagery-only estimates (c) are similar to the imagery+X+FE estimates (d), indicating that pre-treatment satellite features absorb most of the selection-relevant variation that ADM2 fixed effects and standard covariates were proxying.

Relative to prior literature on the World Bank, our image-adjusted estimates are more conservative. Under d, we see clearly positive effects in \emph{Education (110), Transport and Storage (210), General Environmental Protection (410)}, and \emph{Trade and Tourism (330)}. Several other sectors are small or have confidence intervals that include or nearly zero---most notably \emph{Communications (220), Banking and Financial Services (240), Energy Generation and Supply (230), Agriculture, Forestry and Fishing (310),} and especially \emph{Industry, Mining, Construction (320)}. These patterns are broadly consistent with \citet{Bitzer2018} and \citet{Xu2020} but are generally smaller after image adjustment; in contrast to \citet{Bitzer2018}, we do find positive effects for \emph{Education} under most model specifications.

Turning to Chinese aid, results remain large and positive in social sectors such as \emph{Education, Health, Water Supply and Sanitation,} and \emph{Government and Civil Society}, and in economic infrastructure, with \emph{Emergency Response (700)} yielding the largest effects. Smaller image-adjusted effects appear in \emph{Energy Generation and Supply (230)} and \emph{Agriculture, Forestry and Fishing (310)}, which plausibly have longer gestation and slower transmission into the IWI asset bundle.

We next look across all estimation strategies, looking at the maximum and minimum ATE estimates across sector (taking median across strategies). For the World Bank, the maximum (median) ATE is for sector \textit{\WBMaxSector{}}, with a value of \WBMaxSectorATE{}; the minimum  ATE is for sector \textit{\WBMinSector{}}, with a value of \WBMinSectorATE{}. For China, the maximum median ATE is for sector \textit{\CHMaxSector{}}, with a value of \CHMaxSectorATE{}; the minimum is for sector \textit{\CHMinSector{}}, with a value of \CHMinSectorATE{}.

These sector extremes align with differences in time-to-impact and what the asset-based outcome captures. For the World Bank, \textit{Trade \& Tourism (330)} plausibly operates through market‑access and service‑sector channels that can translate relatively quickly into higher earnings and durable goods \citep{WorldBank2008,Donaldson2016}. By contrast, sectors like \textit{\WBMinSector{}} or \textit{\CHMinSector{}} might entail long build‑out cycles or require significant preparatory investments before effects materialize, yielding smaller near‑term IWI gains. On the China side, large \textit{Emergency Response (700)} estimates are consistent with rapid reconstruction, injecting resources and public works \citep{CavalloNoy2011}. These interpretations are descriptive and do not rule out targeting dynamics or residual confounding. 

It is also worth briefly discussing the few consistently negative estimated treatment effects---for example, World Bank's \textit{\WBMinSector{}} interventions sector just discussed. Possibly, these negative or near-zero estimates might reflect how such projects may target weaker or more institutionally fragile areas, where aid is harder to translate into short-term asset gains, so our conservative image-adjusted models capture selection into more challenging environments rather than genuine harm. That said, the within-neighborhood fixed-effects robustness check in Section~\ref{s:unitFE_robustness} yields similarly small (and sometimes negative) point estimates with wide confidence intervals for these two sectors, suggesting more limited within-unit evidence of short-run asset gains. These cases merit further investigation.

\clearpage 

\subsection{Explaining Variation in ATE Estimates}

To summarize how different modeling ingredients affect the estimates, we pool all funder--sector results and regress the \emph{ATE level} on simple indicators for whether a given specification includes tabular covariates ($X$), ADM2 fixed effects (FE), and satellite imagery ($M$), allowing separate intercepts by funder. This provides a compact diagnostic of the extent to which the EO–ML component contributes beyond conventional tabular adjustment. Table \ref{tab:RegATEDiffAnalysis_SEanalytical} reports two parsimonious models. We report both because Model 1 provides an interpretable marginal‑effects benchmark for $X$, FE, and $M$, while Model 2 relaxes additivity by estimating distinct combination effects (e.g., $M{+}X{+}FE$), serving as a robustness check that the imagery‑driven shrinkage in ATEs persists.

Several patterns emerge. First, specifications that incorporate imagery ($M$) produce often \emph{smaller} ATEs than otherwise-comparable models that omit imagery. This is the strongest and most stable association in the table and is consistent with our central claim: images proxy the map-based information donors use when allocating projects, improving confounding control and shrinking observational effects toward zero. Second, adding ADM2 fixed effects (FE) is associated with slightly lower absolute ATEs on average (see robustness section for an individual-level FE analysis), whereas including $X$ alone has little systematic association with the ATE once other ingredients are accounted for. Third, these relationships hold when we allow mean differences by funder; the China mean ATE exceeds the World Bank mean ATE in the pooled regressions, echoing the sector-level results above.

Overall fit is high (adjusted $R^2$ is \AdjRsqModelTwo{} and \AdjRsqModelOne{}), indicating that these simple model ingredients explain most of the across-specification variation. Any model variant containing $M$ (e.g., $M$, $M{+}X$, $M{+}\!FE$, $M{+}X{+}\!FE$) sits below the tabular-only benchmark, reinforcing that the downward shift is broad-based rather than driven by a single specification. In short, imagery-based controls consistently make the estimates more conservative, which is exactly what we would expect if they are capturing selection-relevant features that are missing from standard tabular sets.

\singlespacing

\begin{table}[htbp] \centering 
\footnotesize 
\begin{tabular}{@{\extracolsep{5pt}} lcc} 
\\[-1.8ex]\hline 
\hline \\[-1.8ex] 
 & Model 1 & Model 2 \\ 
\hline \\[-1.8ex] 
Funder: China & 13.39 (13.22)$^*$  & 15.66 (11.55)$^*$  \\ 
Funder: WB & 6.74 (8.46)$^*$  & 9.05 (7.11)$^*$  \\ 
  &  &  \\ 
Contains X & -1.55 (-2.15)$^*$  &  \\ 
Contains FE & -1.15 (-1.59) &  \\ 
Contains M & -2.75 (-3.84)$^*$  &  \\ 
  &  &  \\ 
X &  & -5.08 (-3.03)$^*$  \\ 
FE &  & -4.92 (-3.25)$^*$  \\ 
X+FE &  & -4.72 (-2.82)$^*$  \\ 
M &  & -6.89 (-4.60)$^*$  \\ 
M + X &  & -6.05 (-3.80)$^*$  \\ 
M + FE &  & -5.35 (-3.42)$^*$  \\ 
M + X +FE &  & -7.30 (-4.85)$^*$  \\ 
  &  &  \\ 
\emph{Other statistics} &  &  \\ 
Observations & 200 & 200 \\ 
Adjusted R-squared & 0.71 & 0.73 \\ 
Outcome & ATE & ATE \\ 
\hline \\[-1.8ex] 
\end{tabular} 
  \caption{Outcome: Absolute ATE estimates. Estimator: Weighted OLS (inverse variance weights). $t$-values in parentheses, $*$ indicates $p<0.05$. } 
  \label{tab:RegATEDiffAnalysis_SEanalytical} 
\end{table} 

\defaultspacing

\subsection{Insights about Aid Assignment Mechanisms}

\begin{figure}[htb]
  \centering
\includegraphics[width=1.1\linewidth]{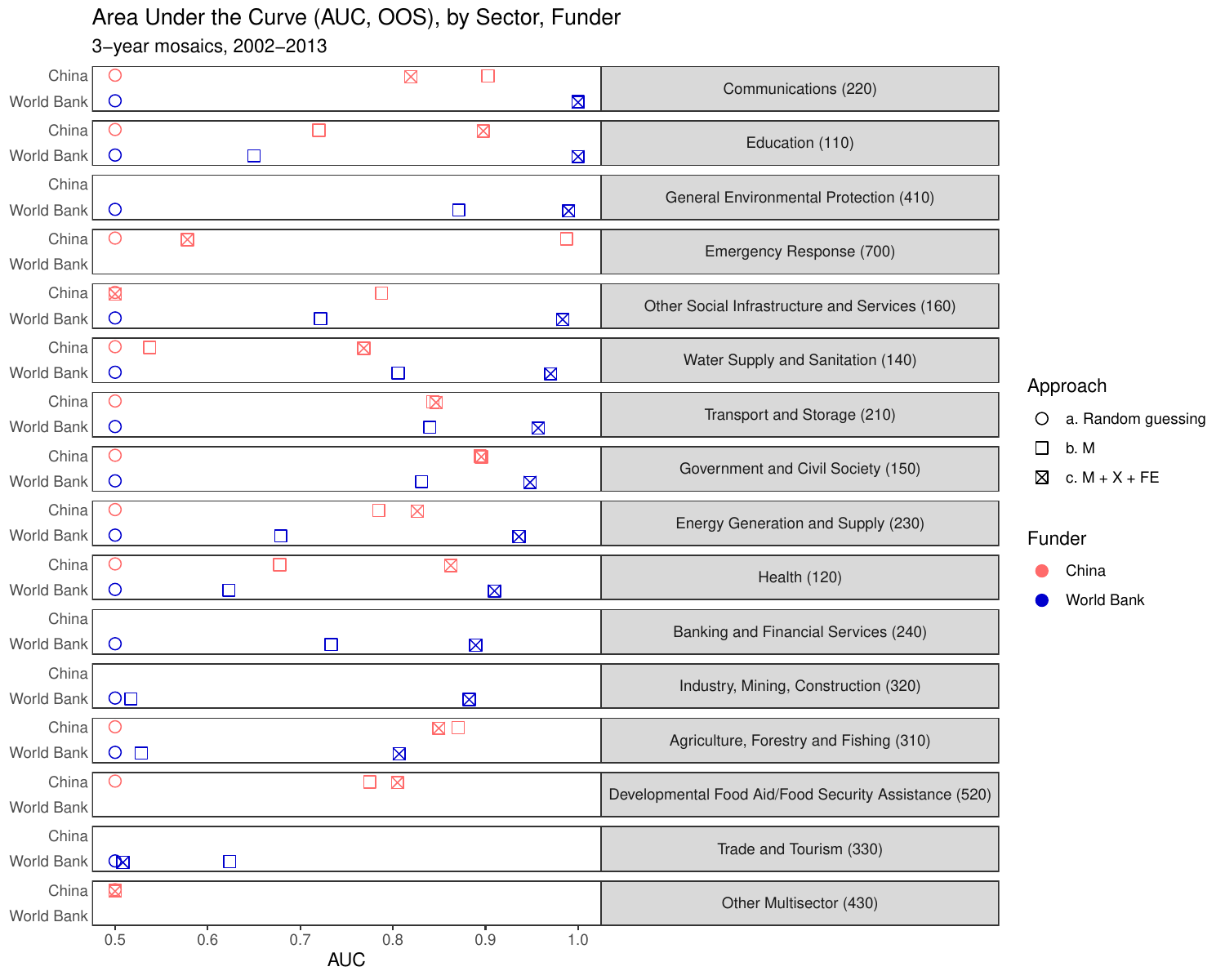}
  \caption{
    Out-of-sample Area Under the Curve (AUC), a metric for evaluating model prediction quality, for each funder, sector, and analysis backbone. Values closer to one are better.  Here, FE refers to ADM2 fixed effects. Out-of-sample observations come from ADM2 areas not used in model training. 
}
  \label{fig:AUC}
\end{figure}

Having assessed the robustness of our impact estimates, we now turn to the assignment mechanism, using out-of-sample prediction performance to quantify how systematically each donor targets neighborhoods. Figure~\ref{fig:AUC} summarizes out‑of‑sample Area Under the Curves (AUC) for predicting which neighborhoods receive projects. AUC values of 0.50 indicate predictive performance no better than random guessing; 0.70 means that, in 70\% of randomly drawn treated--control pairs, the model ranks the actually treated neighborhood higher; values closer to 1 indicate highly predictable placement, based on the data and modeling strategy.

Two broad patterns emerge. First, placement is meaningfully \emph{non‑random} for both funders: models that use satellite imagery alone (M) already separate treated from control units well above the random‑guessing baseline in most cases, and adding tabular covariates with ADM2 fixed effects (M{+}X{+}FE) yields small, consistent improvements. This supports our core identification strategy—image features proxy the map‑based information officials use when deciding where to place projects.

Second, there is sectoral heterogeneity in treatment predictability. Projects with physical requirements show often high levels of predictability using image covariates (e.g., mining), while some social sectors show more moderate separability (Education (110) is an exception). This is the type of variation we would expect if imagery and tabular variables together capture on‑the‑ground constraints and opportunities that are highly non-uniform across settings.

\begin{table}[htbp] \centering 
\footnotesize 
\begin{tabular}{@{\extracolsep{5pt}} lccc} 
\\[-1.8ex]\hline 
\hline \\[-1.8ex] 
 & Model 1 & Model 2 & Model 3 \\ 
\hline \\[-1.8ex] 
WB Funder & 0.08 (2.63)$^*$  &  & 0.08 (2.40)$^*$  \\ 
log(N) &  & 0.17 (4.85)$^*$  & 0.17 (4.81)$^*$  \\ 
Class imbalance &  & 0.40 (2.36)$^*$  & 0.61 (2.97)$^*$  \\ 
  &  &  &  \\ 
\emph{Other statistics} &  &  &  \\ 
Intercept & 0.77 (30.42)$^*$  & -1.23 (-2.75)$^*$  & -1.38 (-3.04)$^*$  \\ 
Observations & 175 & 175 & 175 \\ 
Adjusted R-squared & 0.03 & 0.18 & 0.20 \\ 
Outcome & AUC & AUC & AUC \\ 
\hline \\[-1.8ex] 
\end{tabular} 
  \caption{Outcome: AUC. Baseline funder: China. 
          Estimator: OLS. $t$-values in parentheses, $*$ indicates $p<$0.05. } 
  \label{tab:RegAUCDiffAnalysis_SEanalytical} 
\end{table}

Table~\ref{tab:RegAUCDiffAnalysis_SEanalytical} helps quantify apparent differences between funders. In simple comparisons, World Bank sectors modestly look more predictable than Chinese sectors. Overall, variability in AUC values is largely compositional, being driven by data volume and treatment/control balance, and possibly somewhat intrinsic differences in assignment opacity between World Bank and Chinese funders. 

\clearpage 

\paragraph{Tabular Variable Salience.} 

Figure \ref{fig:SALIENCETAB} in the Appendix summarizes the relationship between tabular covariates and project placement in the tabular–only model ($X{+}FE$) and the image–augmented model ($M{+}X{+}FE$). As our modeling strategy involves non-linear operations (e.g., Transformers), we cannot simply look at model parameters, as we can with linear models; rather, we need to compute the sensitivity of the model's predicted treatment probability to changes in each tabular covariate, averaged across all observations, using automatic differentiation (a computational process that efficiently computes the exact derivatives of complex functions by applying the chain rule to a sequence of elementary arithmetic operations). 

\begin{figure}[htb]
  \centering
  \includegraphics[width=1.\linewidth]{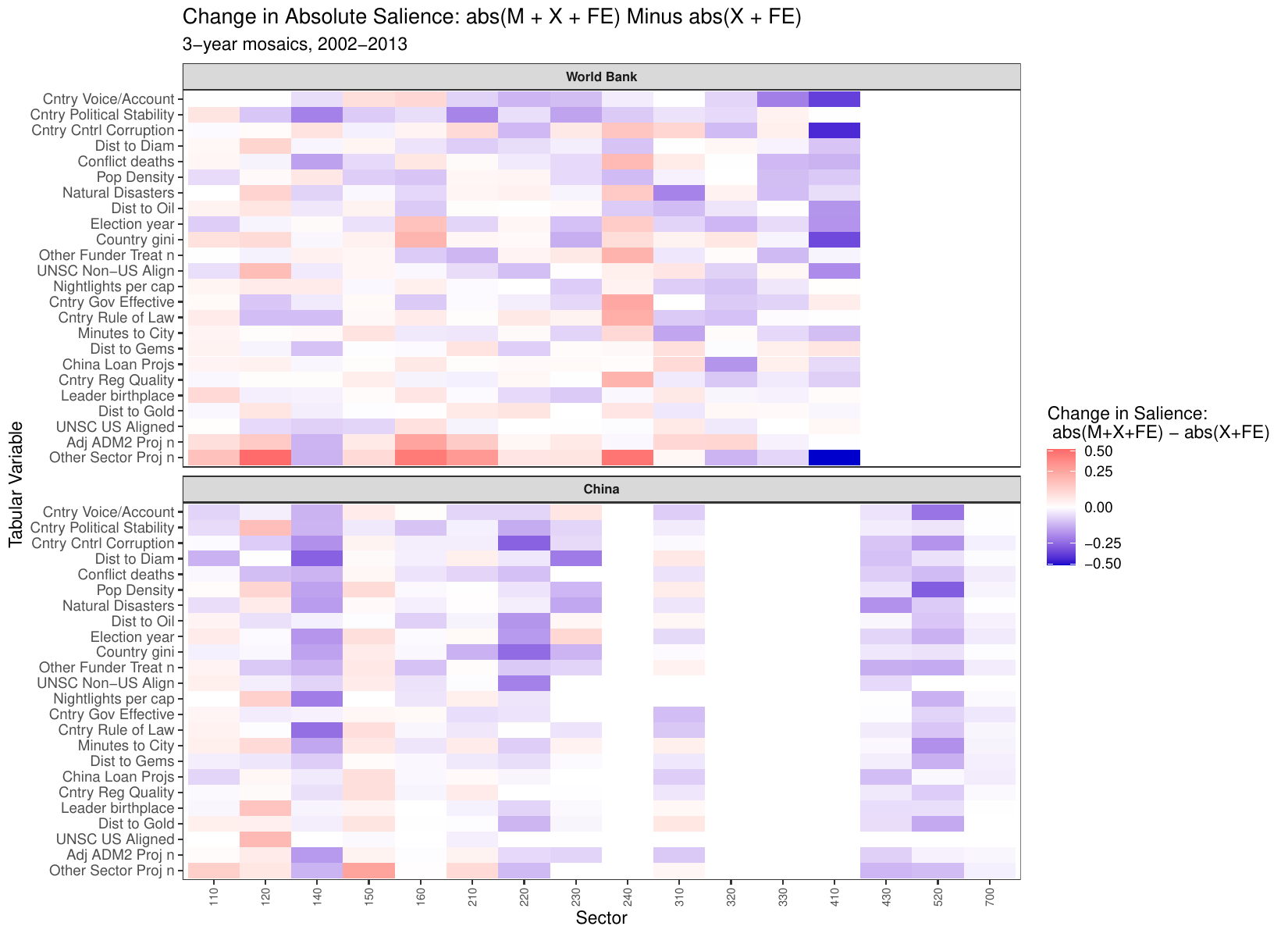}
  \caption{
Change in the absolute salience of tabular variables attributable to tabular confounders, before and after conditioning on image information, across sectors for World Bank and China. Red indicates increased salience magnitude when images are included; blue indicates decreased salience magnitude, and white indicates minimal change. Some columns are left white where there are not enough units to perform analysis for that funder-sector combination.
  }
\label{fig:SALIENCEDELTA}
\end{figure}

In these salience figures the $y$‑axis lists the tabular variables (one row per covariate) and the $x$‑axis lists sectors. Each cell reports a summary of how much a given covariate moves the model’s predicted probability of treatment for that funder–sector. In Figure \ref{fig:SALIENCETAB} (Appendix), the values are \emph{signed} sensitivities: positive values mean that increasing the covariate is associated with a higher predicted chance of a project, negative values mean the opposite. In Figure \ref{fig:SALIENCEDELTA} (here), each cell shows the \emph{change in absolute salience} when images are added, $\lvert \textrm{Salience}(M{+}X{+}FE)\rvert - \lvert \textrm{Salience}(X{+}FE)\rvert$, so \textcolor{red}{red} cells indicate that a variable becomes more influential once images are included, \textcolor{blue}{blue} cells indicate it matters less, and white indicates little change (or too few units for that funder–sector column). Because inputs are standardized before modeling, salience magnitudes are on a comparable scale across variables.

We highlight several observations. 

\emph{First}, adding images often shrinks the absolute salience of many widely used spatial proxies—population density, market access (minutes to city), and country–level governance scores—especially for the World Bank (blue panels in Figure \ref{fig:SALIENCEDELTA} are common). This is consistent with the identification claim: imagery carries geospatial structure that standard tabular sets only approximate, so once we condition on images, those variables matter less.

\emph{Second}, changes for China are more heterogeneous across sectors. In several funder–sector cells, event- or politics‑style variables (e.g., natural disasters, leader birthplace, or UNSC voting) become relatively \emph{more} salient once images absorb geography (red cells in Figure \ref{fig:SALIENCEDELTA}). Qualitatively, this suggests that after we control for what is visible from space, the model reallocates attention toward non‑visual drivers of Chinese project placement in specific sectors.

\emph{Third}, images mainly re-weight variables rather than flip their signs: the direction of association visible in Figure \ref{fig:SALIENCETAB} is largely stable before vs.\ after imaging, but magnitudes compress. Substantively, that means imagery helps separate ``where projects can feasibly go" (geography and the built environment) from ``why they go there" (political and programmatic factors).

To gauge cross‑funder similarity in these salience profiles, we perform  Canonical Correlation Analysis (CCA) on the sector–by–covariate salience matrices from the image‑augmented model. CCA takes two multivariate datasets measured on the same features---in our case, for each sector we have a vector of saliences over tabular variables for the World Bank and another for China---and finds pairs of linear combinations (one per funder) that are maximally correlated across sectors. We treat sectors as observations and covariate saliences as features (columns are standardized so that scale differences do not drive the results). The \emph{leading canonical correlation} summarizes how similar the two donors’ \emph{multivariate} emphasis patterns are across sectors: values near 1 imply that the donors load on similar mixtures of covariates; values near 0 imply largely orthogonal emphasis. This method is appropriate here because it compares full profiles without forcing us to pick one covariate at a time, it is invariant to linear rescaling, and it handles collinearity by working with linear combinations rather than individual variables. The leading canonical correlation here is \CrossSectorCanonicalXSalience{}, indicating limited alignment between Chinese and World Bank placement logics once geography is accounted for. In short: the two donors emphasize different combinations of factors across sectors.

We do not interpret these salience scores causally; they are descriptive of the assignment mechanism under each adjustment set. Their main value here is to show \emph{why} image adjustment reduces ATEs estimates: images soak up spatial structure that tabular proxies were previously carrying, yielding more conservative (and, we argue, more credible) effect estimates.

\subsection{Robustness and Alternative Identification Strategies}\label{s:Robustness}

The main estimates in Figure~\ref{fig:ATEs} rely on selection-on-observables after conditioning on pre-treatment imagery and tabular covariates. We run four complementary robustness checks that probe whether the qualitative conclusions depend on (i) identifying variation only within neighborhoods over time, (ii) including lagged nightlights in the covariate set, (iii) the spatial scale of the IWI outcome grid (motivated by DHS coordinate displacement), and (iv) our treatment-footprint rules for imprecisely geocoded projects. Detailed figures and tables are reported in the Results Appendix.

\begin{figure}[htb]
  \centering
\includegraphics[width=0.77\linewidth]{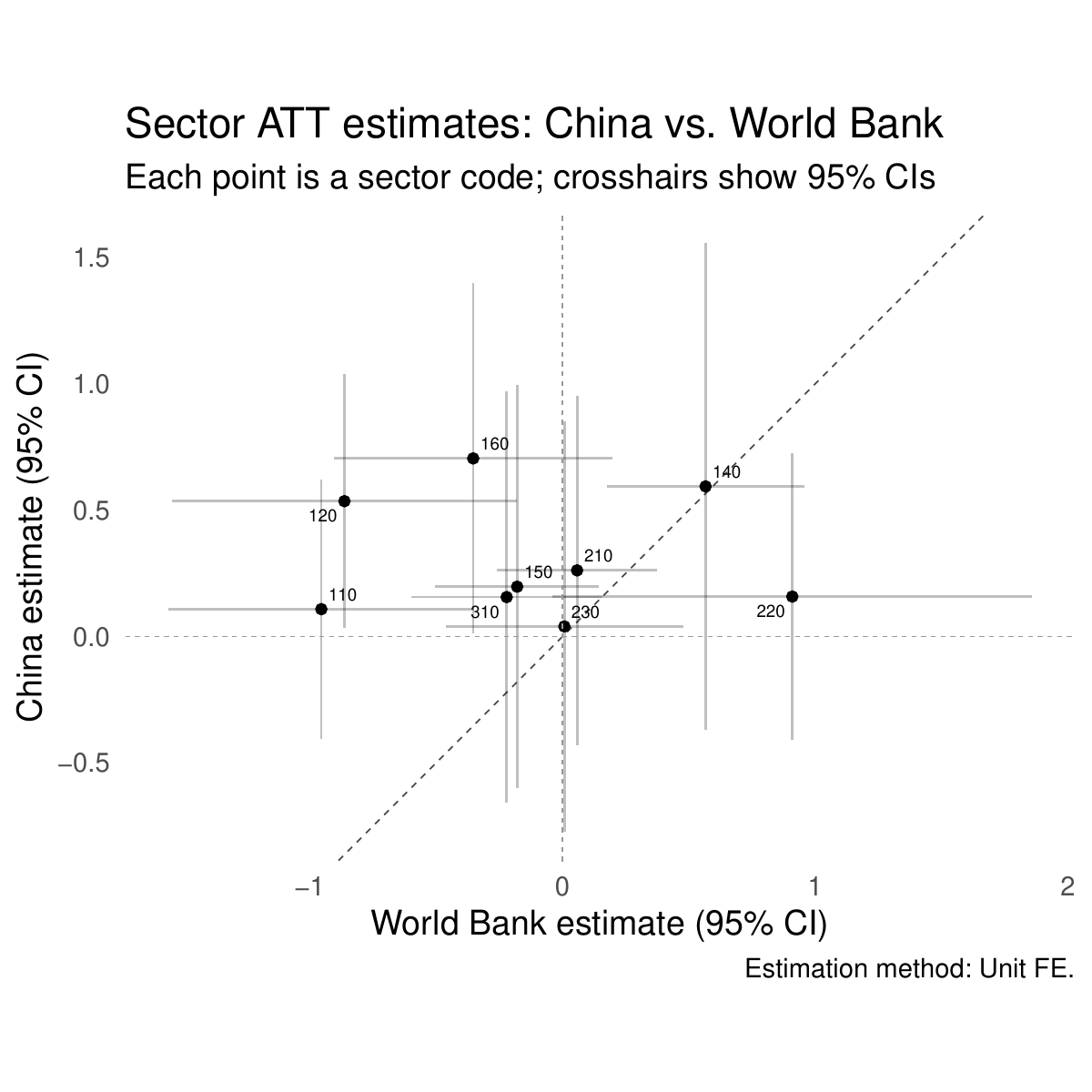}
  \caption{
Unit-level fixed effects results.
}
  \label{fig:UnitFE}
\end{figure}

\paragraph{Within-neighborhood (unit) fixed effects.}\label{s:unitFE_robustness}
To reduce reliance on selection on observables, we estimate a two-way fixed-effects model that uses only within-neighborhood changes in exposure among treatment switchers:
\[
Y_{it}=\alpha_i+\lambda_t+\beta A_{it}+\varepsilon_{it},
\]
where $Y_{it}$ is IWI (measured one period after commitment), $A_{it}$ indicates sector--funder treatment in period $t$, $\alpha_i$ are neighborhood fixed effects, and $\lambda_t$ are period fixed effects. Standard errors are clustered at the ADM2 level. Across Chinese projects, point estimates are directionally consistent with the IPW results ($>0$) but attenuated in magnitude and less precise (see Figure \ref{fig:UnitFE}), reflecting the limited number of switchers. World Bank effects more often switch signs in this unit-level fixed effects analysis. For the World Bank sector with the most consistently negative point estimates (\emph{\WBMinSector{}}), the corresponding fixed-effects estimate is closer to zero and imprecise (\WBMinSectorUnitFEATE{}, SE = \WBMinSectorUnitFEATESE{}). We interpret these results cautiously, pending a more rigorous investigation into switcher dynamics and within-sector heterogeneity that would complicate causal interpretation. 

\paragraph{Robustness to lagged nightlights in the adjustment set.}\label{s:NTLrobustness}
Because lagged per-capita nightlights are persistent and also enter the IWI construction pipeline, we re-estimate the main \ImModel{} M+X+FE specifications after removing lagged nightlights from $\mathbf{X}_{i(t-1)}$. Estimates remain qualitatively consistent: across the $\RobustNCellsCommon{}$ funder--sector cells, the median absolute ATE change is \RobustNTLMedianAbsDiff{} IWI points and the maximum absolute change is \RobustNTLMaxAbsDiff{}. 
The correlation between baseline and no-nightlights ATEs is \RobustNTLCorrATE{}.

\paragraph{Robustness to neighborhood size.}\label{s:raster_robustness}
To assess sensitivity to the 6.7 km $\times$ 6.7 km outcome definition---and the possibility that DHS displacement moves some clusters toward neighborhood edges---we construct a coarser IWI outcome by averaging the four adjacent 6.7 km tiles around each cluster (effective 13.4 km neighborhoods) and re-estimate the main models. ATEs show high but slightly lower correlation with the baseline specification (correlation: \RobustLargeBuffCorrATE{}) compared to the nightlight robustness; the qualitative ordering of effects across funders and sectors is broadly unchanged.

\paragraph{Robustness to treatment definition.}\label{s:strictPrec_robustness}
Finally, we tighten the treatment definition by excluding ADM2-wide assignments that arise from imprecisely geocoded (precision-3) projects and restricting treatment to precision-1 and precision-2 locations. Figure~\ref{fig:ATE_StrictPrec} shows that effects remain broadly consistent with the main analysis; baseline and strict-precision ATEs are correlated (correlation: \RobustStrictPrecCorrATE{}); see the Appendix. That said, World Development estimates are notably higher in magnitude in this regime.

Overall, these robustness tests suggest that conclusions from the main analysis about the relative ranking of sector effects and the relative effects of Chinese vs. World Bank programs are not generally sensitive to modeling choices about nightlights, outcome resolution, or treatment-definition rules. Choice of treatment and especially identification strategy have greatest consequence for the magnitude and precision of sector-specific estimates. Broadly speaking, across identification strategy and treatment/outcome definition, Chinese programs are still estimated here to have greater effects on living conditions. 

\section{Discussion}\label{s:Discussion}

Researchers continually seek to improve the causal understanding of the impact of local development projects on individuals' health and living conditions \citep{imbens2015causal}. In the case of the World Bank and China, there is a lack of localized tabular data on the geographical and historical maps that officials used to strategically decide on where to allocate their projects across Africa in the past two decades \citep{bedi2007more}. Our study applies and extends a recently proposed EO–ML method that proxies these maps with the help of satellite images \cite{Jerzak2023a,Jerzak2023}. The contribution here is less about inventing a new estimator and more about demonstrating, in a large‑scale applied setting, how image‑based confounder adjustment changes estimated impacts and what it reveals about assignment mechanisms. Another innovation here is that we use a novel subnational dataset on health and living conditions recently produced by \cite{Pettersson2023}. Combining these two advances, this study analyzes the effect of locally targeted World Bank and Chinese development projects. The study includes a sector-specific analysis, yielding findings roughly comparable to those of the subnational studies. Our use of nearly 10,000 neighborhoods as potential treated and control observations for each funder and sector analysis enables individual-sector-level analysis, making visible significant sector heterogeneity. Our findings are also robust to dropping lagged nightlights from the covariate set and to substituting alternative proxies for baseline economic conditions, alleviating concerns about mechanically re‑using nightlights in both outcome construction and confounder adjustment. These results advance the aid effectiveness debate by addressing some of the challenges facing subnational development studies. Next, we discuss potential implications. 

\paragraph{Hypothesizing Potential Mechanisms Driving the Results: Why and Where Beneficial Effects?}

A finding of this analysis is that, on average, Chinese aid programs demonstrate a larger positive impact on our machine-learning-derived wealth index than World Bank programs across nearly many comparable sectors (see Figure \ref{fig:ATEs}). This divergence in effect size can likely be attributed to the differing mechanisms through which each donor’s project portfolio and implementation style influence local economies. The literature notes that Chinese aid is heavily concentrated in physical infrastructure sectors, such as transportation, energy, and construction \citep{brautigam2009dragon,Dreher2018a}. 

The mechanism for these projects is often direct: construction creates local, albeit often temporary, employment, injecting cash into the community. Upon completion, infrastructure such as roads and bridges can reduce transportation costs and better connect communities to larger markets, leading to a more widespread diffusion of economic activity \citep{Bluhm2020}. This enhanced connectivity can boost agricultural incomes, create new opportunities for small businesses, and increase labor mobility, all of which generate household income that can then be used to purchase assets and make home improvements, as captured by the International Wealth Index (IWI). While this pathway explains the large effects in sectors like ``Transport and Storage,'' our study also finds significant positive effects for Chinese projects in social sectors such as ``Health'' and ``Education'' \citep{woods2008whose}. This may be due to China's implementation model, which often bypasses recipient government systems in a ``demand-driven'' process that may result in the faster delivery of tangible goods and services \citep{martuscelli2020economics}.

In contrast, the World Bank’s projects yield smaller, though still broadly positive and often statistically significant, effects on the IWI. The Bank's official strategy involves working through recipient government systems to build long-term capacity, a process codified by the Paris Declaration on Aid Effectiveness \citep{OECD2005}. The causal mechanisms for projects in sectors like ``Government and Civil Society'' or ``Other Social Infrastructure'' are therefore more indirect. They aim to improve the quality and accountability of public services, effects that are often slow to materialize and may not translate into measurable household wealth within our study's timeframe. Even for a sector with a direct link to the IWI, like ``Water Supply and Sanitation,'' the World Bank’s process-oriented approach may be slower than China’s more direct implementation \citep{isaksson2018chinese}. The weaker performance of World Bank aid in ``Industry, Mining, Construction'' and ``Agriculture,'' as found in our analysis, further suggests its model may be less suited to generating the kind of immediate, capital-intensive impact that our wealth measure captures effectively \citep{ssozi2019effectiveness}.

This distinction between China's more direct, often faster implementation style and the World Bank's more indirect, institution-building approach offers a compelling explanation for our findings. The ``hard'' and ``fast'' nature of Chinese aid appears to generate more immediate and tangible improvements in material living conditions as measured by satellite imagery and the IWI. The World Bank's approach, while potentially laying the groundwork for more sustainable, long-term development, produces smaller and less immediate wealth effects. Our study's ability to detect this heterogeneity underscores that the ``how'' of development aid is just as critical as the ``how much'' \citep{easterly2003can}.

Additionally, \citet{Pettersson2023} wealth estimates overcome measurement limitations in nightlight data used to proxy subnational wealth, overcome the lack of panel data in Demographic and Health Surveys, and enable the use of fixed effects at the ADM2 (county, district, city) level to control for unobserved, time-invariant confounders in subnational regions. 

Moreover, our use of satellite-image deconfounding methods for causal inference yielded overall more conservative estimates compared to tabular methods (with fixed effects) only. This empirical evidence suggests that our EO-ML method effectively adjusts for confounding information not available in existing tabular sources. Consequently, this implies that previous studies are likely overestimating the beneficial effects of development projects when they fail to adjust for image features. 

The image-based confounding estimation method demonstrated a wide range of AUC model quality scores in predicting treatment probability across funders, sectors, and modeling strategies. We expected image-based confounding to be more beneficial for sectors dependent on the physical conditions of treatment areas compared to those that can be placed anywhere (e.g., Energy Generation vs. Education), but did not find a clear pattern meeting those expectations.

Data resolution is a concern in studies using satellite methods and data. If the resolution of satellite images is not sufficiently high, the algorithm may miss confounding factors.  The multi-resolution nature of our data raises several challenges to statistical inference (see \citet{Meng2014}), including issues related to mixing data measured at multiple resolutions and to estimating an estimand at a finer-grained resolution (\SpatialRes{} km neighborhoods) than many of the covariates upon which it is based.  
To ensure the use of nationally representative samples in each country included in each funder/sector analysis, we use locations of neighborhoods surveyed in one round of Demographic and Health Surveys as units of analysis.  We use \citet{Pettersson2023}'s
wealth estimates for those locations over time.  This is a non-random sample, since all these countries have a relationship with the U.S. Future research could use wealth estimates to create broader samples of neighborhoods across the African continent, including in countries never surveyed by DHS, or could include all human settlement areas across Africa instead of a sample. With the possible dismantling of funds to USAID, and its detrimental effect on future DHS data collection efforts, EO-ML methods such as \citet{Pettersson2023} will be even more critical for aid evaluation.

\paragraph{Future Work.} There are several promising angles that future research can build on. For example,  future research can replicate this study incorporating novel debiasing methods in the handling of ML-generated outcomes \citep{angelopoulos2023prediction,petterssonDebiasingMachineLearning2025,risterportinarimarancaCorrectingMeasurementErrors2025}, uncertainty quantification in the imputed-image-derived outcomes \cite{kakooei2024increasing}, evaluating the effect of satellite image scale \cite{pmlr-v275-zhu25a}, and  incorporating large-language models for enhanced interpretability \cite{murugaboopathyPlatonicRepresentationsPoverty2025}. 

The fact that project effects spill over into adjacent ADM2s violates the non-interference requirement of the stable unit treatment value assumption (SUTVA) and creates an ongoing challenge for this and other studies.  Future work can also employ techniques (e.g., excluding neighborhoods adjacent to treated ADM2s from the control group) to account for spatial interdependence. Alternatively, studies can explicitly model the impact of spatio-network interference \citep{reich2021review}.

In conclusion, this study contributes to the ongoing discussion on the impact of local development projects and their effectiveness by demonstrating how EO-ML methods can innovate in addressing causal challenges in development research. \hfill $\square$

\paragraph{Data and code availability.} All code to reproduce the analyses in this paper, along with replication datasets derived from publicly available sources, will be made available at 
\begin{itemize}
\item[] \url{https://doi.org/10.7910/DVN/XYI1PG} 
\end{itemize}
and 
\begin{itemize}
\item[] \url{https://github.com/AIandGlobalDevelopmentLab/} 
\end{itemize}

\newpage
\bibliographystyle{plainnat} 
\bibliography{mybib}

\begin{thebibliography}{115}
\providecommand{\natexlab}[1]{#1}
\providecommand{\url}[1]{\texttt{#1}}
\expandafter\ifx\csname urlstyle\endcsname\relax
  \providecommand{\doi}[1]{doi: #1}\else
  \providecommand{\doi}{doi: \begingroup \urlstyle{rm}\Url}\fi

\bibitem[Ahmed(2022)]{Ahmed2022}
S~Ahmed.
\newblock {China's Official Finance in the Global South: What's the Literature
  Telling Us?}
\newblock \emph{Review of Economics}, 73\penalty0 (3):\penalty0 223--252, 2022.
\newblock URL \url{https://doi.org/10.1515/roe-2021-0030}.

\bibitem[AidData(2017)]{AidData2017}
AidData.
\newblock {World Bank Geocoded Research Release, Version 1.4.2 Geocoded Dataset
  [Technical Report]}, 2017.
\newblock URL
  \url{https://www.aiddata.org/data/world-bank-geocoded-research-release-level-1-v1-4-2}.

\bibitem[Andersen(2006)]{Andersen2006}
Hansen H. \& Markussen~T Andersen, T.~B.
\newblock {US politics and World Bank IDA-lending}.
\newblock \emph{The Journal of Development Studies}, 42\penalty0 (5):\penalty0
  772--794, 2006.
\newblock URL \url{https://doi.org/10.1080/00220380600741946}.

\bibitem[Angelopoulos et~al.(2023)Angelopoulos, Bates, Fannjiang, Jordan, and
  Zrnic]{angelopoulos2023prediction}
Anastasios~N Angelopoulos, Stephen Bates, Clara Fannjiang, Michael~I Jordan,
  and Tijana Zrnic.
\newblock {Prediction-powered Inference}.
\newblock \emph{Science}, 382\penalty0 (6671):\penalty0 669--674, 2023.

\bibitem[Asatullaeva(2021)]{Asatullaeva2021}
Aghdam R. F. Z. Ahmad N. \& Tashpulatova~L Asatullaeva, Z.
\newblock {The Impact of Foreign Aid on Economic Development: A Systematic
  Literature Review and Content Analysis of the Top 50 Most Influential
  Papers}.
\newblock \emph{Journal of International Development}, 33:\penalty0 717--751,
  2021.
\newblock URL \url{https://doi.org/10.1002/jid.3543}.

\bibitem[Banerjee et~al.(2015)Banerjee, Duflo, Glennerster, and
  Kinnan]{Banerjee2015}
Abhijit Banerjee, Esther Duflo, Rachel Glennerster, and Cynthia Kinnan.
\newblock {The Miracle of Microfinance? Evidence from a Randomized Evaluation}.
\newblock \emph{American Economic Journal: Applied Economics}, 7\penalty0
  (1):\penalty0 22--53, 2015.

\bibitem[Bank(2023)]{GiniCoefficient2023}
World Bank.
\newblock {Gini Index}, 2023.
\newblock URL \url{https://data.worldbank.org/indicator/SI.POV.GINI}.

\bibitem[Bedi et~al.(2007{\natexlab{a}})Bedi, Coudouel, and
  Simler]{bedi2007more}
Tara Bedi, Aline Coudouel, and Kenneth Simler.
\newblock \emph{{More Than a Pretty Picture: Using Poverty Maps to Design
  Better Policies and Interventions}}.
\newblock World Bank Publications, 2007{\natexlab{a}}.

\bibitem[Bedi et~al.(2007{\natexlab{b}})Bedi, Coudouel, and
  Simler]{bediMorePrettyPicture2007}
Tara Bedi, Aline Coudouel, and Kenneth Simler.
\newblock More {{Than}} a {{Pretty Picture}} : {{Using Poverty Maps}} to
  {{Design Better Policies}} and {{Interventions}}.
\newblock Technical report, World Bank, Washington, DC, 2007{\natexlab{b}}.

\bibitem[BenYishay(2022)]{BenYishay2022}
DiLorenzo M. \& Dolan~C BenYishay, A.
\newblock {The Economic Efficiency of Aid Targeting}.
\newblock \emph{World Development}, 160\penalty0 (106062), 2022.
\newblock URL \url{https://doi.org/10.1016/j.worlddev.2022.106062}.

\bibitem[Berlin et~al.(2023)Berlin, Desai, and Olofsg{\aa}rd]{Berlin2023}
Maria~Perrotta Berlin, Raj~M Desai, and Anders Olofsg{\aa}rd.
\newblock {Trading Favors? UN Security Council Membership and Subnational
  Favoritism in Aid Recipients}.
\newblock \emph{The Review of International Organizations}, 18\penalty0
  (2):\penalty0 237--258, 2023.

\bibitem[Bitzer and G{\"o}ren(2018)]{Bitzer2018}
J{\"u}rgen Bitzer and Erkan G{\"o}ren.
\newblock {Foreign Aid and Subnational Development: A Grid Cell Analysis}.
\newblock Technical report, Oldenburg Discussion Papers in Economics, 2018.
\newblock URL \url{https://www.econstor.eu/handle/10419/175419}.

\bibitem[Bluhm(2020)]{Bluhm2020}
Dreher A. Fuchs A. Parks B. Strange A. \& Tierney M.~J Bluhm, R.
\newblock {Connective Financing: Chinese Infrastructure Projects and the
  Diffusion of Economic Activity in Developing Countries}, 2020.
\newblock URL \url{https://doi.org/10.2139/ssrn.3262101}.

\bibitem[Bomprezzi et~al.(2024)Bomprezzi, Dreher, Fuchs, Hailer, Kammerlander,
  Kaplan, Marchesi, Masi, Robert, and Unfried]{bomprezzi2024wedded}
Pietro Bomprezzi, Axel Dreher, Andreas Fuchs, Teresa Hailer, Andreas
  Kammerlander, Lennart~C Kaplan, Silvia Marchesi, Tania Masi, Charlotte
  Robert, and Kerstin Unfried.
\newblock {Wedded to Prosperity? Informal Influence and Regional Favoritism}.
\newblock 2024.

\bibitem[Brautigam(2009)]{brautigam2009dragon}
Deborah Brautigam.
\newblock \emph{{The Dragon's Gift: The Real Story of China in Africa}}.
\newblock Oxford University Press, 2009.

\bibitem[Brazys et~al.(2021)Brazys, Vadlamannati, and Dietrich]{Brazys2021}
Samuel Brazys, Krishna~Chaitanya Vadlamannati, and Simone Dietrich.
\newblock {Does Chinese Development Finance Raise Household Welfare in Low- and
  Middle-income Countries?}
\newblock \emph{World Development}, 140:\penalty0 105345, 2021.
\newblock \doi{10.1016/j.worlddev.2020.105345}.

\bibitem[Briggs(2021)]{Briggs2021}
R.~C Briggs.
\newblock {Why Does Aid Not Target the Poorest?}
\newblock \emph{International Studies Quarterly}, 65\penalty0 (3):\penalty0
  739--752, 2021.
\newblock URL \url{https://doi.org/10.1093/isq/sqab035}.

\bibitem[Briggs(2017)]{Briggs2017}
Ryan~C. Briggs.
\newblock {Does Foreign Aid Target the Poorest?}
\newblock \emph{International Organization}, 71\penalty0 (1):\penalty0
  187--206, 2017.
\newblock \doi{10.1017/S0020818316000345}.
\newblock URL \url{https://doi.org/10.1017/S0020818316000345}.

\bibitem[Broich(2017)]{Broich2017}
Tobias Broich.
\newblock {Do Authoritarian Regimes Receive More Chinese Development Finance
  than Democratic Ones? Empirical Evidence for Africa}.
\newblock \emph{China Economic Review}, 46:\penalty0 180--207, 2017.
\newblock URL \url{https://doi.org/10.1016/j.chieco.2017.09.006}.

\bibitem[Burgert et~al.(2013)Burgert, Zachary, and Colston]{Burgert2013}
Clara~R Burgert, Blake Zachary, and Josh Colston.
\newblock {Incorporating Geographic Information into Demographic and Health
  Surveys: A Field Guide to GPs Data Collection}.
\newblock \emph{Fairfax: ICF International}, 2013.
\newblock URL
  \url{https://dhsprogram.com/publications/publication-dhsm9-dhs-questionnaires-and-manuals.cfm}.

\bibitem[Burke et~al.(2021)Burke, Driscoll, Lobell, and
  Ermon]{burkeUsingSatelliteImagery2021}
Marshall Burke, Anne Driscoll, David~B. Lobell, and Stefano Ermon.
\newblock {Using Satellite Imagery to Understand and Promote Sustainable
  Development}.
\newblock \emph{Science}, 371\penalty0 (6535):\penalty0 eabe8628, March 2021.
\newblock ISSN 0036-8075, 1095-9203.
\newblock \doi{10.1126/science.abe8628}.

\bibitem[Cavallo and Noy(2009)]{CavalloNoy2011}
Eduardo Cavallo and Ilan Noy.
\newblock {The Economics of Natural Disasters: a Survey}.
\newblock Technical report, IDB working paper series, 2009.

\bibitem[Chai and Tang(2023)]{chai2023world}
Qingyuan Chai and Zhongyi Tang.
\newblock {The World Bank and China: Comparing the Impacts of Their Development
  Projects in Africa}.
\newblock \emph{Available at SSRN 4598476}, 2023.

\bibitem[Chernozhukov et~al.(2018)Chernozhukov, Chetverikov, Demirer, Duflo,
  Hansen, Newey, and Robins]{chernozhukov2018double}
Victor Chernozhukov, Denis Chetverikov, Mert Demirer, Esther Duflo, Christian
  Hansen, Whitney Newey, and James Robins.
\newblock {Double/debiased Machine Learning for Treatment and Structural
  Parameters}, 2018.

\bibitem[Chin(2012)]{Chin2012}
Gregory~T Chin.
\newblock {China as a ‘Net Donor’: Tracking Dollars and Sense}.
\newblock \emph{Cambridge Review of International Affairs}, 25\penalty0
  (4):\penalty0 579--603, 2012.

\bibitem[Chong et~al.(2009)Chong, Gradstein, and Calderon]{chong2009can}
Alberto Chong, Mark Gradstein, and Cecilia Calderon.
\newblock {Can Foreign Aid Reduce Income Inequality and Poverty?}
\newblock \emph{Public Choice}, 140\penalty0 (1):\penalty0 59--84, 2009.

\bibitem[Cinelli et~al.(2024)Cinelli, Forney, and Pearl]{cinelli2024crash}
Carlos Cinelli, Andrew Forney, and Judea Pearl.
\newblock {A Crash Course in Good and Bad Controls}.
\newblock \emph{Sociological Methods \& Research}, 53\penalty0 (3):\penalty0
  1071--1104, 2024.

\bibitem[Croft et~al.(2018)Croft, Marshall, and Allen]{Croft2018}
Tom~N. Croft, Amber M.~J. Marshall, and Courtney~K. Allen.
\newblock \emph{{Guide to DHS Statistics}}.
\newblock ICF, 2018.
\newblock URL
  \url{https://dhsprogram.com/data/Guide-to-DHS-Statistics/index.cfm}.
\newblock DHS Program.

\bibitem[Croicu and Sundberg(2015)]{Croicu2015}
Mihai Croicu and Ralph Sundberg.
\newblock {UCDP Georeferenced Event Dataset Codebook Version 4.0}.
\newblock \emph{Journal of Peace Research}, 50\penalty0 (4):\penalty0 523--532,
  2015.

\bibitem[Daoud and Dubhashi(2023)]{daoud_statistical_2023}
Adel Daoud and Devdatt Dubhashi.
\newblock Statistical {Modeling}: {The} {Three} {Cultures}.
\newblock \emph{Harvard Data Science Review}, 5\penalty0 (1), January 2023.
\newblock ISSN 2644-2353, 2688-8513.
\newblock \doi{10.1162/99608f92.89f6fe66}.
\newblock URL \url{https://hdsr.mitpress.mit.edu/pub/uo4hjcx6/release/1}.

\bibitem[Daoud and Johansson()]{daoudImpactAusterityChildren2024}
Adel Daoud and Fredrik~D. Johansson.
\newblock {The Impact of Austerity on Children: {{Uncovering}} Effect
  Heterogeneity by Political, Economic, and Family Factors in Low- and
  Middle-Income Countries}.
\newblock 118:\penalty0 102973.
\newblock ISSN 0049-089X.
\newblock \doi{10.1016/j.ssresearch.2023.102973}.
\newblock URL
  \url{https://www.sciencedirect.com/science/article/pii/S0049089X2300128X}.

\bibitem[Daoud et~al.(2017)Daoud, Nosrati, Reinsberg, Kentikelenis, Stubbs, and
  King]{daoud_impact_2017}
Adel Daoud, Elias Nosrati, Bernhard Reinsberg, Alexander~E. Kentikelenis,
  Thomas~H. Stubbs, and Lawrence~P. King.
\newblock Impact of {International} {Monetary} {Fund} programs on child health.
\newblock \emph{Proceedings of the National Academy of Sciences}, 114\penalty0
  (25):\penalty0 6492--6497, May 2017.
\newblock ISSN 0027-8424, 1091-6490.
\newblock \doi{10.1073/pnas.1617353114}.
\newblock URL \url{http://www.pnas.org/content/early/2017/05/09/1617353114}.
\newblock 00002.

\bibitem[Daoud et~al.(2021)Daoud, Jordan, Sharma, Johansson, Dubhashi, Paul,
  and Banerjee]{Daoud2022}
Adel Daoud, Felipe Jordan, Makkunda Sharma, Fredrik Johansson, Devdatt
  Dubhashi, Sourabh Paul, and Subhashis Banerjee.
\newblock Using satellites and artificial intelligence to measure health and
  material-living standards in india.
\newblock \emph{arXiv preprint arXiv:2202.00109}, 2021.

\bibitem[Daoud et~al.(2022)Daoud, Herlitz, and
  Subramanian]{daoudIMFFairnessCalibrating2022}
Adel Daoud, Anders Herlitz, and S.~V. Subramanian.
\newblock {{{IMF}} Fairness: {{Calibrating}} the Policies of the
  {{International Monetary Fund}} Based on Distributive Justice}.
\newblock \emph{World Development}, 157:\penalty0 105924, September 2022.
\newblock ISSN 0305-750X.
\newblock \doi{10.1016/j.worlddev.2022.105924}.

\bibitem[Delforge(2023)]{Delforge2023}
Wathelet V. Below R. Sofia C. L. Tonnelier M. van Loenhout J. \& Speybroeck~N
  Delforge, D.
\newblock {EM-DAT: the Emergency Events Database}, 2023.

\bibitem[Demir and Duan(2024)]{Demir2023}
Firat Demir and Yi~Duan.
\newblock {Target at the Right Level: Aid, Spillovers, and Growth in
  Sub-saharan Africa}.
\newblock \emph{Applied Economics}, 56\penalty0 (28):\penalty0 3293--3333,
  2024.

\bibitem[Dolan and McDade(2020)]{Dolan2020}
Carrie~B Dolan and Kaci~Kennedy McDade.
\newblock {Pulling the Purse Strings: Are There Sectoral Differences in
  Political Preferencing of Chinese Aid to Africa?}
\newblock \emph{PloS one}, 15\penalty0 (4):\penalty0 e0232126, 2020.

\bibitem[Donaldson and Storeygard(2016)]{Donaldson2016}
Dave Donaldson and Adam Storeygard.
\newblock {The View from Above: Applications of Satellite Data in Economics}.
\newblock \emph{Journal of Economic Perspectives}, 30\penalty0 (4):\penalty0
  171--198, 2016.

\bibitem[Drabo(2021)]{Drabo2021}
Alassane Drabo.
\newblock \emph{{How Do Climate Shocks Affect the Impact of FDI, ODA and
  Remittances on Economic Growth?}}
\newblock International Monetary Fund, 2021.

\bibitem[Dreher and Fuchs(2015)]{Dreher2015}
Axel Dreher and Andreas Fuchs.
\newblock {Rogue Aid? An Empirical Analysis of China's Aid Allocation}.
\newblock \emph{Canadian Journal of Economics}, 48\penalty0 (3):\penalty0
  988--1023, 2015.

\bibitem[Dreher and Lohmann(2015)]{Dreher2015a}
Axel Dreher and Steffen Lohmann.
\newblock {Aid and Growth at the Regional Level}.
\newblock \emph{Oxford Review of Economic Policy}, 31\penalty0 (3-4):\penalty0
  420--446, 2015.

\bibitem[Dreher et~al.(2009)Dreher, Sturm, and Vreeland]{Dreher2009}
Axel Dreher, Jan-Egbert Sturm, and James~Raymond Vreeland.
\newblock {Development Aid and International Politics: Does Membership on the
  UN Security Council Influence World Bank Decisions?}
\newblock \emph{Journal of Development Economics}, 88\penalty0 (1):\penalty0
  1--18, 2009.

\bibitem[Dreher et~al.(2018)Dreher, Fuchs, Parks, Strange, and
  Tierney]{Dreher2018a}
Axel Dreher, Andreas Fuchs, Brad Parks, Austin~M Strange, and Michael~J
  Tierney.
\newblock {Apples and Dragon Fruits: The Determinants of Aid and Other Forms of
  State Financing from China to Africa}.
\newblock \emph{International Studies Quarterly}, 62\penalty0 (1):\penalty0
  182--194, 02 2018.
\newblock ISSN 0020-8833.
\newblock \doi{10.1093/isq/sqx052}.
\newblock URL \url{https://doi.org/10.1093/isq/sqx052}.

\bibitem[Dreher et~al.(2021)Dreher, Fuchs, Parks, Strange, and
  Tierney]{Dreher2021b}
Axel Dreher, Andreas Fuchs, Bradley Parks, Austin Strange, and Michael~J
  Tierney.
\newblock Aid, china, and growth: Evidence from a new global development
  finance dataset.
\newblock \emph{American Economic Journal: Economic Policy}, 13\penalty0
  (2):\penalty0 135--174, 2021.

\bibitem[Dreher et~al.(2022{\natexlab{a}})Dreher, Fuchs, Parks, Strange, and
  Tierney]{Dreher2022}
Axel Dreher, Andreas Fuchs, Bradley Parks, Austin Strange, and Michael~J
  Tierney.
\newblock \emph{{Banking on Beijing: The Aims and Impacts of China's Overseas
  Development Program}}.
\newblock Cambridge University Press, 2022{\natexlab{a}}.
\newblock URL \url{https://doi.org/10.1017/9781108564496}.

\bibitem[Dreher et~al.(2022{\natexlab{b}})Dreher, Lang, Rosendorff, and
  Vreeland]{Dreher2022a}
Axel Dreher, Valentin Lang, B~Peter Rosendorff, and James~Raymond Vreeland.
\newblock {Bilateral or Multilateral? International Financial Flows and the
  Dirty-work Hypothesis}.
\newblock \emph{The Journal of Politics}, 84\penalty0 (4):\penalty0 1932--1946,
  2022{\natexlab{b}}.

\bibitem[Dreher(2019)]{Dreher2019}
Fuchs A. Hodler R. Parks B. C. Raschky P. A. \& Tierney M.~J Dreher, A.
\newblock {African Leaders and the Geography of China's Foreign Assistance}.
\newblock \emph{Journal of Development Economics}, 140, 2019.
\newblock URL \url{https://doi.org/10.1016/j.jdeveco.2019.04.003}.

\bibitem[Duflo(2015)]{Duflo2015}
Greenstone M. Guiteras R. \& Clasen~T Duflo, E.
\newblock {Toilets Can Work: Short and Medium Run Health Impacts of Addressing
  Complementarities and Externalities in Water and Sanitation (Working Paper
  21521)}.
\newblock \emph{National Bureau of Economic Research}, 2015.
\newblock URL \url{https://doi.org/10.3386/w21521}.

\bibitem[D’Amour et~al.(2021)D’Amour, Ding, Feller, Lei, and
  Sekhon]{DAmour2021}
Alexander D’Amour, Peng Ding, Avi Feller, Lihua Lei, and Jasjeet Sekhon.
\newblock {Overlap in Observational Studies with High-Dimensional Covariates}.
\newblock \emph{Journal of Econometrics}, 221\penalty0 (2):\penalty0 644--654,
  2021.

\bibitem[Easterly(2003)]{easterly2003can}
William Easterly.
\newblock {Can Foreign Aid Buy Growth?}
\newblock \emph{Journal of Economic Perspectives}, 17\penalty0 (3):\penalty0
  23--48, 2003.

\bibitem[Easterly(2006)]{Easterly2005}
William Easterly.
\newblock \emph{{The White Man's Burden: Why the West's Efforts to Aid the Rest
  Have Done So Much Ill and So Little Good}}.
\newblock Penguin Press, 2006.

\bibitem[Francazi et~al.(2023)Francazi, Baity-Jesi, and
  Lucchi]{francazi2023theoretical}
Emanuele Francazi, Marco Baity-Jesi, and Aurelien Lucchi.
\newblock {A Theoretical Analysis of the Learning Dynamics under Class
  Imbalance}.
\newblock In \emph{International Conference on Machine Learning}, pages
  10285--10322. PMLR, 2023.

\bibitem[Freedman and Berk(2008)]{freedman2008weighting}
David~A Freedman and Richard~A Berk.
\newblock {Weighting Regressions by Propensity Scores}.
\newblock \emph{Evaluation Review}, 32\penalty0 (4):\penalty0 392--409, 2008.

\bibitem[Galiani et~al.(2005)Galiani, Gertler, and Schargrodsky]{Galiani2005}
Sebastian Galiani, Paul Gertler, and Ernesto Schargrodsky.
\newblock {The Causal Effect of Water Quality on Diarrheal Diseases in
  Children}.
\newblock \emph{Econometrica}, 73\penalty0 (5):\penalty0 1505--1548, 2005.
\newblock \doi{10.1111/j.1468-0262.2005.00631.x}.

\bibitem[Gehring(2022)]{Gehring2022}
Kaplan L. C. \& Wong M. H.~L Gehring, K.
\newblock {China and the World Bank---How Contrasting Development Approaches
  Affect the Stability of African States}, 2022.
\newblock URL \url{https://doi.org/10.1016/j.jdeveco.2022.102902}.

\bibitem[Gilmore(2005)]{Gilmore2005}
Gleditsch N. P. Lujala P. \& R\o{}d J.~K Gilmore, E.
\newblock {Conflict Diamonds: A New Dataset}, 2005.

\bibitem[Gre{\ss}er and Stadelmann(2021)]{Greer2021a}
Christina Gre{\ss}er and David Stadelmann.
\newblock {Evaluating Water-and Health-related Development Projects: A
  Cross-project and Micro-based Approach}.
\newblock \emph{The Journal of Development Studies}, 57\penalty0 (7):\penalty0
  1221--1239, 2021.

\bibitem[Guillon and Mathonnat(2020)]{Guillon2020}
Marl{\`e}ne Guillon and Jacky Mathonnat.
\newblock {What Can We Learn on Chinese Aid Allocation Motivations from
  Available Data? A Sectorial Analysis of Chinese Aid to African Countries}.
\newblock \emph{China Economic Review}, 60:\penalty0 101265, 2020.

\bibitem[Halleröd et~al.(2013)Halleröd, Rothstein, Daoud, and
  Nandy]{hallerod_bad_2013}
Björn Halleröd, Bo~Rothstein, Adel Daoud, and Shailen Nandy.
\newblock Bad governance and poor children: a comparative analysis of
  government efficiency and severe child deprivation in 68 low-and
  middle-income countries.
\newblock \emph{World Development}, 48:\penalty0 19--31, 2013.
\newblock URL
  \url{http://www.sciencedirect.com/science/article/pii/S0305750X13000831}.

\bibitem[Huang et~al.(2016)Huang, Sun, Liu, Sedra, and
  Weinberger]{huang2016deep}
Gao Huang, Yu~Sun, Zhuang Liu, Daniel Sedra, and Kilian~Q Weinberger.
\newblock {Deep Networks with Stochastic Depth}.
\newblock In \emph{European Conference on Computer Vision}, pages 646--661.
  Springer, 2016.

\bibitem[Imbens and Rubin(2015)]{imbens2015causal}
Guido~W. Imbens and Donald~B. Rubin.
\newblock \emph{{Causal Inference for Statistics, Social, and Biomedical
  Sciences: An Introduction}}.
\newblock Cambridge University Press, 2015.
\newblock \doi{10.1017/CBO9781139025751}.

\bibitem[Isaksson and Kotsadam(2018)]{isaksson2018chinese}
Ann-Sofie Isaksson and Andreas Kotsadam.
\newblock {Chinese Aid and Local Corruption}.
\newblock \emph{Journal of Public Economics}, 159:\penalty0 146--159, 2018.

\bibitem[Jean(2016)]{Jean2016}
Burke M. Xie M. Davis W. M. Lobell D. B. \& Ermon~S Jean, N.
\newblock Combining satellite imagery and machine learning to predict poverty.
  science, 353(6301), 790--794, 2016.
\newblock URL \url{https://doi.org/10.1126/science.aaf7894}.

\bibitem[Jerzak(2023{\natexlab{a}})]{Jerzak2023}
\&~Daoud~A Jerzak, Connor~T.
\newblock {CausalImages: An R Package for Causal Inference with Earth
  Observation, Bio-medical, and Social Science Images (arXiv:2310.00233).
  arXiv}, 2023{\natexlab{a}}.
\newblock URL \url{https://doi.org/10.48550/arXiv.2310.00233}.

\bibitem[Jerzak et~al.(2023)Jerzak, Johansson, and Daoud]{jerzak2023image}
Connor~T. Jerzak, Fredrik~Daniel Johansson, and Adel Daoud.
\newblock Image-based treatment effect heterogeneity.
\newblock In \emph{Conference on Causal Learning and Reasoning}, pages
  531--552. PMLR, 2023.

\bibitem[Jerzak(2023{\natexlab{b}})]{Jerzak2023a}
Johansson F. \& Daoud~A Jerzak, C.~T.
\newblock {Integrating Earth Observation Data into Causal Inference: Challenges
  and Opportunities}, 2023{\natexlab{b}}.
\newblock URL \url{https://arxiv.org/abs/2301.12985}.

\bibitem[Kakooei and Daoud(2024)]{kakooei2024increasing}
Mohammad Kakooei and Adel Daoud.
\newblock {Increasing the Confidence of Predictive Uncertainty: Earth
  Observations and Deep Learning for Poverty Estimation}.
\newblock \emph{IEEE Transactions on Geoscience and Remote Sensing},
  62:\penalty0 1--13, 2024.

\bibitem[Kaufmann and Kraay(2024)]{Kaufmann2023}
Daniel Kaufmann and Aart Kraay.
\newblock {The Worldwide Governance Indicators}.
\newblock 2024.

\bibitem[Kim and Kim(2021)]{Kim2021}
Y.~Kim and Y.~Kim.
\newblock {The Autonomy of International Organizations? The Analysis of Major
  Powers’ Influence Over the World Bank’s Aid Policies}.
\newblock \emph{International Area Studies Review}, 24\penalty0 (3):\penalty0
  224--240, 2021.
\newblock \doi{10.1177/22338659211024879}.

\bibitem[Kino et~al.(2021)Kino, Hsu, Shiba, Chien, Mita, Kawachi, and
  Daoud]{kino_scoping_2021}
Shiho Kino, Yu-Tien Hsu, Koichiro Shiba, Yung-Shin Chien, Carol Mita, Ichiro
  Kawachi, and Adel Daoud.
\newblock A scoping review on the use of machine learning in research on social
  determinants of health: {Trends} and research prospects.
\newblock \emph{SSM - Population Health}, 15:\penalty0 100836, September 2021.
\newblock ISSN 2352-8273.
\newblock \doi{10.1016/j.ssmph.2021.100836}.
\newblock URL
  \url{https://www.sciencedirect.com/science/article/pii/S2352827321001117}.

\bibitem[Krueger and Lindahl(2001)]{Krueger2001}
Alan~B. Krueger and Mikael Lindahl.
\newblock {Education for Growth: Why and for Whom?}
\newblock \emph{Journal of Economic Literature}, 39\penalty0 (4):\penalty0
  1101--1136, 2001.
\newblock \doi{10.1257/jel.39.4.1101}.

\bibitem[Lee et~al.(2021)Lee, Park, and Kim]{Lee2021}
JY~Lee, J~Park, and Y~Kim.
\newblock {Chinese Development Finance and Local-Level Inequality in Africa}.
\newblock \emph{The Chinese Journal of International Politics}, 14\penalty0
  (4):\penalty0 533--564, 2021.
\newblock \doi{10.1093/cjip/poab015}.

\bibitem[Leiderer(2021)]{Leiderer2021}
Stefan Leiderer.
\newblock {The Effects of Chinese Aid on Household Welfare in Africa}.
\newblock \emph{World Development}, 140:\penalty0 105221, 2021.
\newblock URL \url{10.1016/j.worlddev.2020.105221}.

\bibitem[Lujala(2007)]{Lujala2007}
R\o{}d J. K. \& Thieme~N Lujala, P.
\newblock {Fighting Over Oil: Introducing a New Dataset}.
\newblock \emph{Conflict Management and Peace Science}, 24\penalty0
  (3):\penalty0 239--256, 2007.

\bibitem[Malik et~al.(2021)Malik, Parks, Russell, Lin, Walsh, Solomon, Zhang,
  Elston, and Goodman]{Malik2021}
Ammar Malik, Bradley Parks, Brooke Russell, Joyce Lin, Katherine Walsh, Kyra
  Solomon, Sheng Zhang, T~Elston, and Seth Goodman.
\newblock {Banking on the Belt and Road: Insights from a New Global Dataset of
  13,427 Chinese Development Projects}.
\newblock \emph{Williamsburg, VA: AidData at William \& Mary}, pages 23--36,
  2021.

\bibitem[Mandon and Woldemichael(2023)]{Mandon2023}
Pierre Mandon and Martha~Tesfaye Woldemichael.
\newblock {Has Chinese Aid Benefited Recipient Countries? Evidence from a
  Meta-regression Analysis}.
\newblock \emph{World Development}, 166:\penalty0 106211, 2023.

\bibitem[Martel(2021)]{Martel2021}
Romane Martel.
\newblock {Can Foreign Aid Reduce Inequality? The Importance of the Domestic
  Context}.
\newblock \emph{The Journal of International Trade \& Economic Development},
  30\penalty0 (7):\penalty0 1028--1053, 2021.
\newblock \doi{10.1080/09638199.2021.1947265}.

\bibitem[Martina(2020)]{Martina2020}
Mirkuz Martina.
\newblock {Does Chinese Aid Improve Health Outcomes? Evidence from a New
  Dataset}.
\newblock \emph{Journal of Development Studies}, 56\penalty0 (5):\penalty0
  943--960, 2020.
\newblock \doi{10.1080/00220388.2019.1627993}.

\bibitem[Martorano et~al.(2020)Martorano, Metzger, and
  Sanfilippo]{martorano2020chinese}
Bruno Martorano, Laura Metzger, and Marco Sanfilippo.
\newblock {Chinese Development Assistance and Household Welfare in Sub-Saharan
  Africa}.
\newblock \emph{World Development}, 129:\penalty0 104909, 2020.
\newblock ISSN 0305-750X.
\newblock \doi{10.1016/j.worlddev.2020.104909}.
\newblock URL
  \url{https://www.sciencedirect.com/science/article/pii/S0305750X20300358}.

\bibitem[Martuscelli(2020)]{martuscelli2020economics}
Antonio Martuscelli.
\newblock {The Economics of China's Engagement with Africa: What Is the
  Empirical Evidence?}
\newblock \emph{Development Policy Review}, 38\penalty0 (3):\penalty0 285--302,
  2020.

\bibitem[McGillivray et~al.(2006)McGillivray, Feeny, Hermes, and
  Lensink]{McGillivray2006}
Mark McGillivray, Simon Feeny, Niels Hermes, and Robert Lensink.
\newblock {Controversies Over the Impact of Development Aid: It Works; It
  Doesn't; It Can, But That Depends}.
\newblock \emph{Journal of International Development: The Journal of the
  Development Studies Association}, 18\penalty0 (7):\penalty0 1031--1050, 2006.

\bibitem[Meng(2014)]{Meng2014}
Xiao-Li Meng.
\newblock {A Trio of Inference Problems That Could Win You a Nobel Prize in
  Statistics (If You Help Fund It)}.
\newblock \emph{Past, Present, and Future of Statistical Science}, pages
  537--562, 2014.
\newblock URL \url{https://doi.org/10.1201/b16720-50}.

\bibitem[Murugaboopathy et~al.(2025)Murugaboopathy, Jerzak, and
  Daoud]{murugaboopathyPlatonicRepresentationsPoverty2025}
Satiyabooshan Murugaboopathy, Connor~T. Jerzak, and Adel Daoud.
\newblock Platonic {{Representations}} for {{Poverty Mapping}}: {{Unified
  Vision-Language Codes}} or {{Agent-Induced Novelty}}?, August 2025.

\bibitem[NASA(2021, November 30)]{NASAnd}
NASA.
\newblock Landsat 5, 2021, November 30.
\newblock URL \url{https://landsat.gsfc.nasa.gov/satellites/landsat-5/}.

\bibitem[OECD(2005)]{OECD2005}
OECD.
\newblock {Paris Declaration on Aid Effectiveness}, 2005.
\newblock URL \url{https://doi.org/10.1787/9789264098084-en}.

\bibitem[{\"O}hler and Nunnenkamp(2014)]{Ohler2014}
Hannes {\"O}hler and Peter Nunnenkamp.
\newblock {Needs-based Targeting or Favoritism? The Regional Allocation of
  Multilateral Aid Within Recipient Countries}.
\newblock \emph{Kyklos}, 67\penalty0 (3):\penalty0 420--446, 2014.

\bibitem[{\"O}hler et~al.(2019){\"O}hler, Negre, Smets, Massari, and
  Bogeti{\'c}]{Ohler2019}
Hannes {\"O}hler, Mario Negre, Lodewijk Smets, Renzo Massari, and {\v{Z}}eljko
  Bogeti{\'c}.
\newblock {Putting Your Money Where Your Mouth Is: Geographic Targeting of
  World Bank Projects to the Bottom 40 Percent}.
\newblock \emph{PloS One}, 14\penalty0 (6):\penalty0 e0218671, 2019.

\bibitem[Pearl(2009)]{pearl2009causality}
Judea Pearl.
\newblock \emph{Causality}.
\newblock Cambridge University Press, 2009.

\bibitem[Pettersson et~al.(2025)Pettersson, Jerzak, and
  Daoud]{petterssonDebiasingMachineLearning2025}
Markus Pettersson, Connor~T. Jerzak, and Adel Daoud.
\newblock Debiasing {{Machine Learning Predictions}} for {{Causal Inference
  Without Additional Ground Truth Data}}: "{{One Map}}, {{Many Trials}}" in
  {{Satellite-Driven Poverty Analysis}}, August 2025.

\bibitem[Pettersson et~al.(2023)Pettersson, Kakooei, Ortheden, Johansson, and
  Daoud]{Pettersson2023}
Markus~B Pettersson, Mohammad Kakooei, Julia Ortheden, Fredrik~D Johansson, and
  Adel Daoud.
\newblock {Time Series of Satellite Imagery Improve Deep Learning Estimates of
  Neighborhood-Level Poverty in Africa}.
\newblock In \emph{IJCAI}, pages 6165--6173, 2023.

\bibitem[Platteau(2004)]{Platteau2004}
J.-P Platteau.
\newblock {Monitoring Elite Capture in Community‑driven Development}.
\newblock \emph{Development and Change}, 35\penalty0 (2):\penalty0 223--246,
  2004.

\bibitem[Prince(2023)]{prince2023understanding}
Simon~J.D. Prince.
\newblock \emph{Understanding Deep Learning}.
\newblock The MIT Press, 2023.
\newblock URL \url{http://udlbook.com}.

\bibitem[Reich et~al.(2021)Reich, Yang, Guan, Giffin, Miller, and
  Rappold]{reich2021review}
Brian~J Reich, Shu Yang, Yawen Guan, Andrew~B Giffin, Matthew~J Miller, and Ana
  Rappold.
\newblock {A Review of Spatial Causal Inference Methods for Environmental and
  Epidemiological Applications}.
\newblock \emph{International Statistical Review}, 89\penalty0 (3):\penalty0
  605--634, 2021.

\bibitem[Research and Unit(2017)]{AidDataResearchandEvaluationUnit2017}
AidData Research and Evaluation Unit.
\newblock {Geocoding Methodology, Version 2.0. AidData at William \& Mary},
  2017.
\newblock URL
  \url{https://www.aiddata.org/publications/geocoding-methodology-version-2-0}.

\bibitem[Rister Portinari~Maranca et~al.()Rister Portinari~Maranca, Chung,
  Hinck, Wolsky, Egami, and
  Stewart]{risterportinarimarancaCorrectingMeasurementErrors2025}
Alessandra Rister Portinari~Maranca, Jihoon Chung, Musashi Hinck, Adam~D.
  Wolsky, Naoki Egami, and Brandon~M. Stewart.
\newblock Correcting the {{Measurement Errors}} of {{AI-Assisted Labeling}} in
  {{Image Analysis Using Design-Based Supervised Learning}}.
\newblock 54\penalty0 (3):\penalty0 984--1016.
\newblock ISSN 0049-1241.
\newblock \doi{10.1177/00491241251333372}.
\newblock URL \url{https://doi.org/10.1177/00491241251333372}.

\bibitem[Rosvold and Buhaug(2021)]{Rosvold2021}
Elisabeth~L Rosvold and Halvard Buhaug.
\newblock {GDIS, A Global Dataset of Geocoded Disaster Locations}.
\newblock \emph{Scientific Data}, 8\penalty0 (1):\penalty0 61, 2021.

\bibitem[Rubin(3)]{Rubin1990}
D.~B Rubin.
\newblock {Formal Model of Statistical Inference for Causal Effects}.
\newblock \emph{Journal of Statistical Planning and Inference}, 25:\penalty0
  279--292, 3.
\newblock URL \url{https://doi.org/10.1016/0378-3758(90)90077-8}.

\bibitem[Rubin(1991)]{rubin1991practical}
Donald~B Rubin.
\newblock {Practical Implications of Modes of Statistical Inference for Causal
  Effects and the Critical Role of the Assignment Mechanism}.
\newblock \emph{Biometrics}, pages 1213--1234, 1991.

\bibitem[Rutstein(2015)]{rutstein2015steps}
Shea~O Rutstein.
\newblock {Steps to Constructing the New DHS Wealth Index}.
\newblock \emph{Rockville, MD: ICF International}, 6, 2015.

\bibitem[Sakamoto et~al.(2025)Sakamoto, Jerzak, and
  Daoud]{sakamotoScopingReviewEarth2025}
Kazuki Sakamoto, Connor~T. Jerzak, and Adel Daoud.
\newblock A {{Scoping Review}} of {{Earth Observation}} and {{Machine
  Learning}} for {{Causal Inference}}: {{Implications}} for the {{Geography}}
  of {{Poverty}}, April 2025.

\bibitem[Sardoschau and Jarotschkin(2024)]{Sardoschau2019}
Sulin Sardoschau and Alexandra Jarotschkin.
\newblock {Chinese Aid in Africa: Attitudes and Conflict}.
\newblock \emph{European Journal of Political Economy}, 81:\penalty0 102500,
  2024.

\bibitem[Science(2021, November 30)]{Landsat7LandsatSciencend}
Landsat 7 |~Landsat Science.
\newblock https://landsat.gsfc, 2021, November 30.
\newblock URL \url{https://landsat.gsfc.}

\bibitem[Scott(2009)]{WorldBank2008}
Allen~J Scott.
\newblock {World Development Report 2009: Reshaping Economic Geography}, 2009.

\bibitem[Skarda(2022)]{Skardand}
Ieva Skarda.
\newblock {Is Income Inequality Important for Foreign Aid Effectiveness?}, sep
  2022.
\newblock URL \url{https://doi.org/10.21203/rs.3.rs-2073414/v1}.
\newblock Preprint.

\bibitem[Smits and Steendijk(2015)]{Smits2015}
Jeroen Smits and Roel Steendijk.
\newblock {The International Wealth Index (IWI)}.
\newblock \emph{Social Indicators Research}, 122\penalty0 (1):\penalty0 65--85,
  2015.

\bibitem[Ssozi et~al.(2019)Ssozi, Asongu, and Amavilah]{ssozi2019effectiveness}
John Ssozi, Simplice Asongu, and Voxi~Heinrich Amavilah.
\newblock {The Effectiveness of Development Aid for Agriculture in Sub-Saharan
  Africa}.
\newblock \emph{Journal of Economic Studies}, 46\penalty0 (2):\penalty0
  284--305, 2019.

\bibitem[Strange(2017)]{Strange2017}
Dreher A. Fuchs A. Parks B. \& Tierney M.~J Strange, A.~M.
\newblock {Tracking Underreported Financial Flows: China's Development Finance
  and the Aid--Conflict Nexus Revisited}, 2017.
\newblock URL \url{https://doi.org/10.1177/0022002715604363}.

\bibitem[Vaswani et~al.(2017)Vaswani, Shazeer, Parmar, Uszkoreit, Jones, Gomez,
  Kaiser, and Polosukhin]{vaswani2017attention}
Ashish Vaswani, Noam Shazeer, Niki Parmar, Jakob Uszkoreit, Llion Jones,
  Aidan~N Gomez, {\L}ukasz Kaiser, and Illia Polosukhin.
\newblock {Attention is All You Need}.
\newblock \emph{Advances in Neural Information Processing Systems}, 30, 2017.

\bibitem[Warmerdam and van Dijk(2013)]{Warmerdam2013}
Ward Warmerdam and Meine~Pieter van Dijk.
\newblock {Chinese State-owned Enterprise Investments in Uganda: Findings from
  a Recent Survey of Chinese Firms in Kampala}.
\newblock \emph{Journal of Chinese Political Science}, 18\penalty0
  (3):\penalty0 281--301, 2013.

\bibitem[Weiss et~al.(2018)Weiss, Nelson, Gibson, Temperley, Peedell, Lieber,
  Hancher, Poyart, Belchior, Fullman, et~al.]{Nelson2008}
D~J Weiss, Andy Nelson, HS~Gibson, W~Temperley, Stephen Peedell, Allie Lieber,
  Matt Hancher, Eduardo Poyart, Sim{\~a}o Belchior, Nancy Fullman, et~al.
\newblock {A Global Map of Travel Time to Cities to Assess Inequalities in
  Accessibility in 2015}.
\newblock \emph{Nature}, 553\penalty0 (7688):\penalty0 333--336, 2018.

\bibitem[Winters(2010)]{Winters2010}
Matthew~S Winters.
\newblock {Choosing to Target: What Types of Countries Get Different Types of
  World Bank Projects}.
\newblock \emph{World Politics}, 62\penalty0 (3):\penalty0 422--458, 2010.

\bibitem[Woods(2008)]{woods2008whose}
Ngaire Woods.
\newblock {Whose Aid? Whose Influence? China, Emerging Donors and the Silent
  Revolution in Development Assistance}.
\newblock \emph{International Affairs}, 84\penalty0 (6):\penalty0 1205--1221,
  2008.

\bibitem[WorldPop(2023)]{WorldPopPopulationDensitynd}
WorldPop.
\newblock {Global High Resolution Population Denominators Project: Population
  Density}, 2023.
\newblock URL \url{https://hub.worldpop.org/geodata/listing?id=77}.
\newblock Funded by The Bill and Melinda Gates Foundation (OPP1134076).
  Retrieved July 24, 2023.

\bibitem[Xu et~al.(2020)Xu, Zhang, and Sun]{Xu2020}
Z.~Xu, Y.~Zhang, and Y.~Sun.
\newblock {Will Foreign Aid Foster Economic development? Grid Panel Data
  Evidence from China’s Aid to Africa}.
\newblock \emph{Emerging Markets Finance and Trade}, 56\penalty0 (14):\penalty0
  3383--3404, 2020.

\bibitem[Zhu et~al.(2025)Zhu, Jerzak, and Daoud]{pmlr-v275-zhu25a}
Fucheng~Warren Zhu, Connor~T. Jerzak, and Adel Daoud.
\newblock Optimizing multi-scale representations to detect effect heterogeneity
  using earth observation and computer vision: Applications to two anti-poverty
  rcts.
\newblock In Biwei Huang and Mathias Drton, editors, \emph{Proceedings of the
  Fourth Conference on Causal Learning and Reasoning}, volume 275 of
  \emph{Proceedings of Machine Learning Research}, pages 894--919. PMLR, 07--09
  May 2025.
\newblock URL \url{https://proceedings.mlr.press/v275/zhu25a.html}.

\end{thebibliography}

\newpage 

\renewcommand\thefigure{A.I.\arabic{figure}} \setcounter{figure}{0}
\renewcommand\thetable{A.I.\arabic{table}} \setcounter{table}{0}

\section{Additional Background Figures} 

\begin{figure}[htb]
  \centering
  \includegraphics[width=0.45\linewidth]{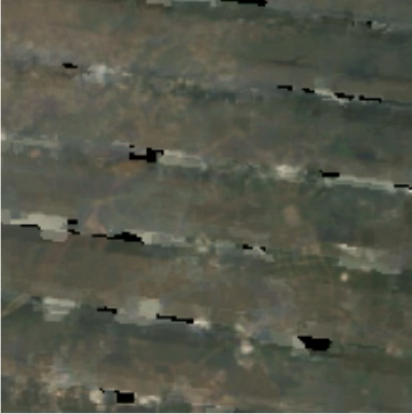}
  \includegraphics[width=0.45\linewidth]{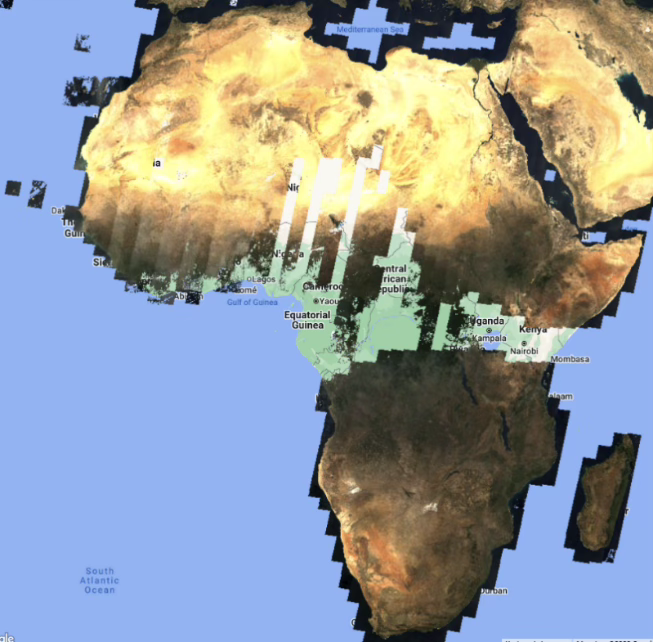}
  \caption{
  Missing data in daytime satellite images.  \textsc{Left:} An image over Cameroon in 2005-2007 with diagonal lines due to the Landsat 7 Scan Line Corrector hardware failure.  Location: latitude 4.014, longitude 9.800. \textsc{Right:} Landsat 5 areas of unsaved data. Source: Author queries in Google Earth Engine.
  }
  \label{fig:Coverage}
\end{figure}

\begin{figure}[htb]
\centering
\begin{subfigure}[b]{0.45\textwidth}
\centering
\includegraphics[width=\textwidth,height=\textwidth,keepaspectratio=true]{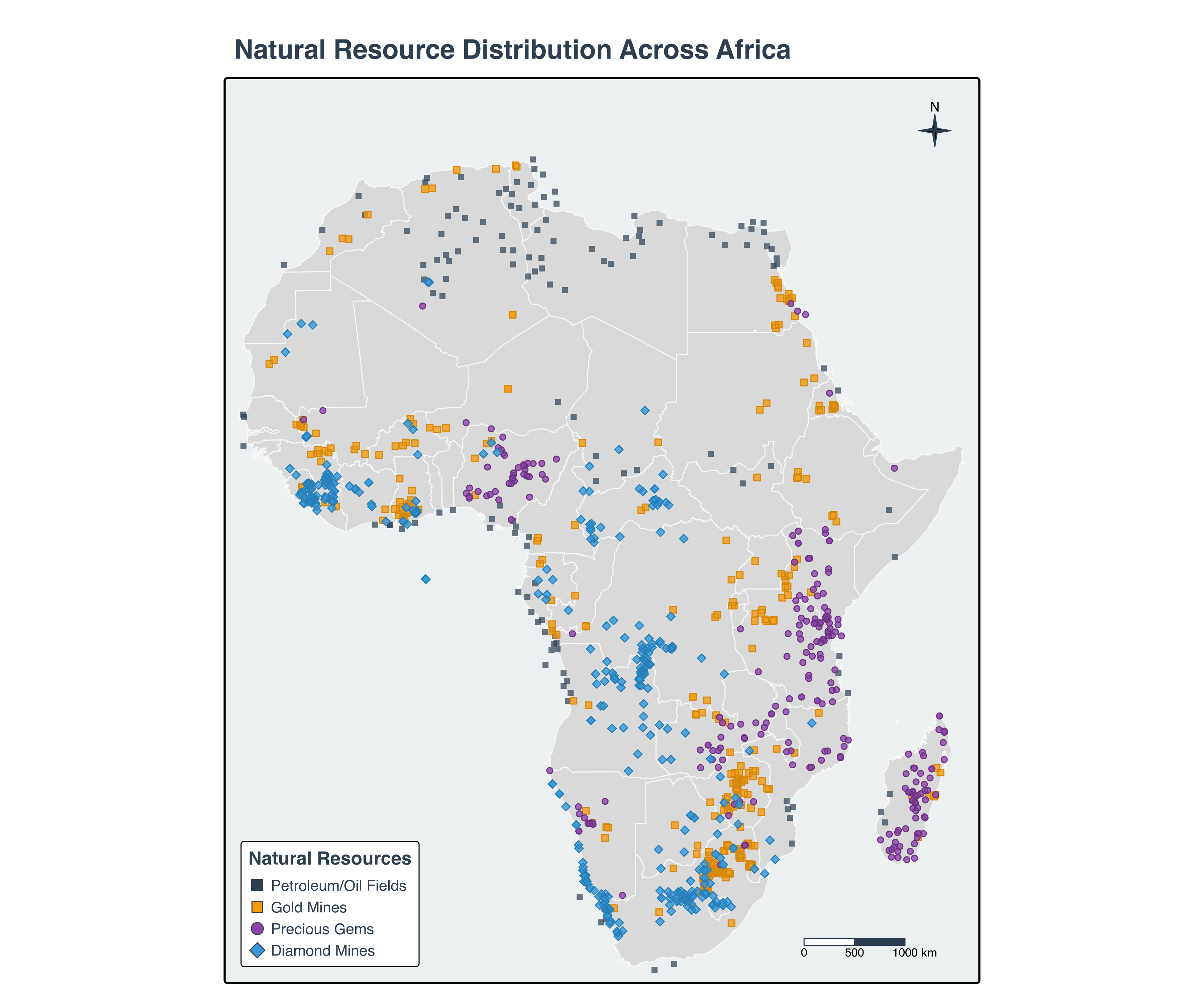}
\caption{Natural resources}
\label{subfig:Resources}
\end{subfigure}
\hfill
\begin{subfigure}[b]{0.45\textwidth}
\centering
\includegraphics[width=\textwidth,height=\textwidth,keepaspectratio=true]{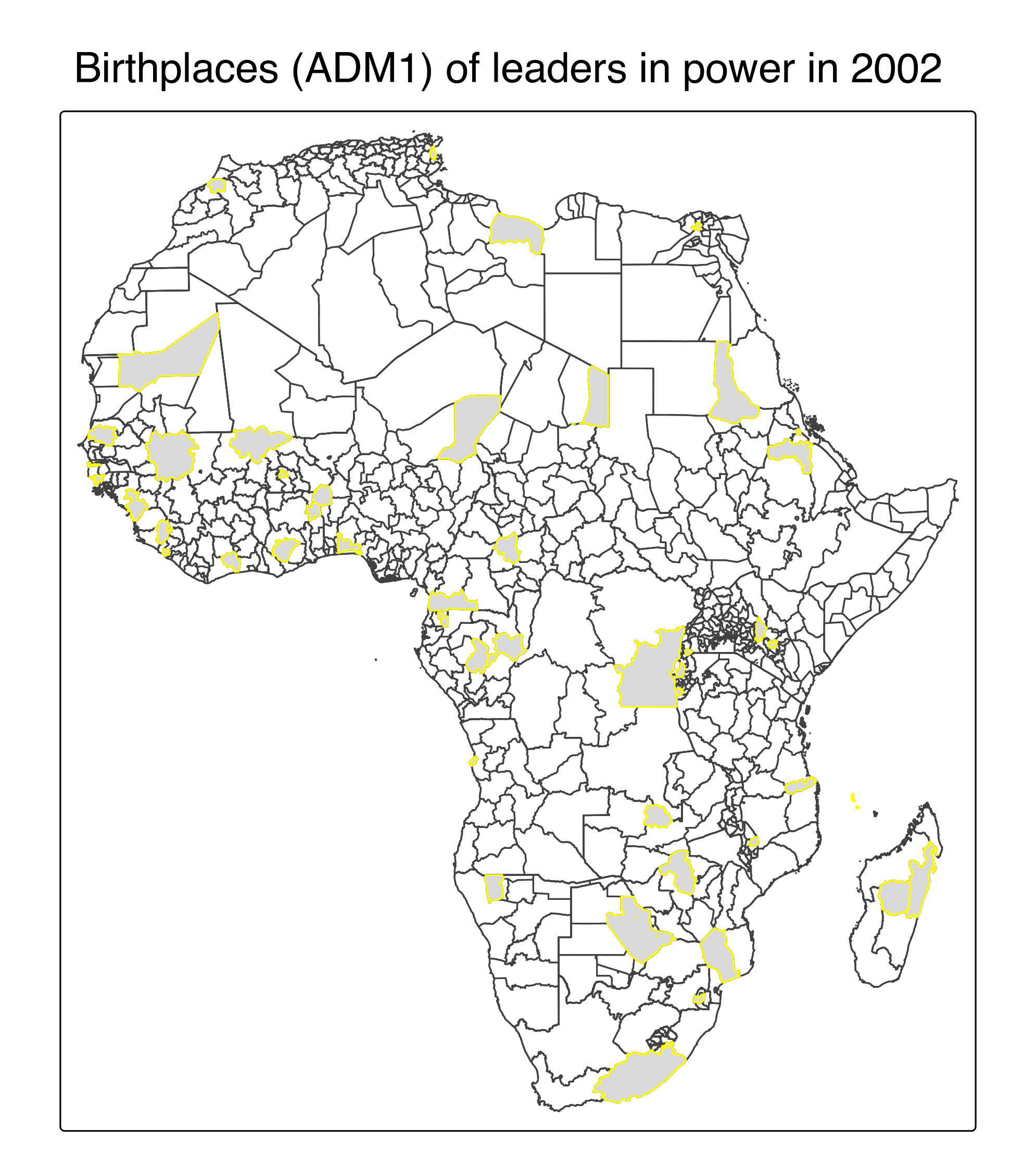}
\caption{Leader birthplaces (2002)}
\label{subfig:PLAD}
\end{subfigure}
\vspace{0.5em}
\begin{subfigure}[b]{0.45\textwidth}
\centering
\includegraphics[width=\textwidth,height=\textwidth,keepaspectratio=true]{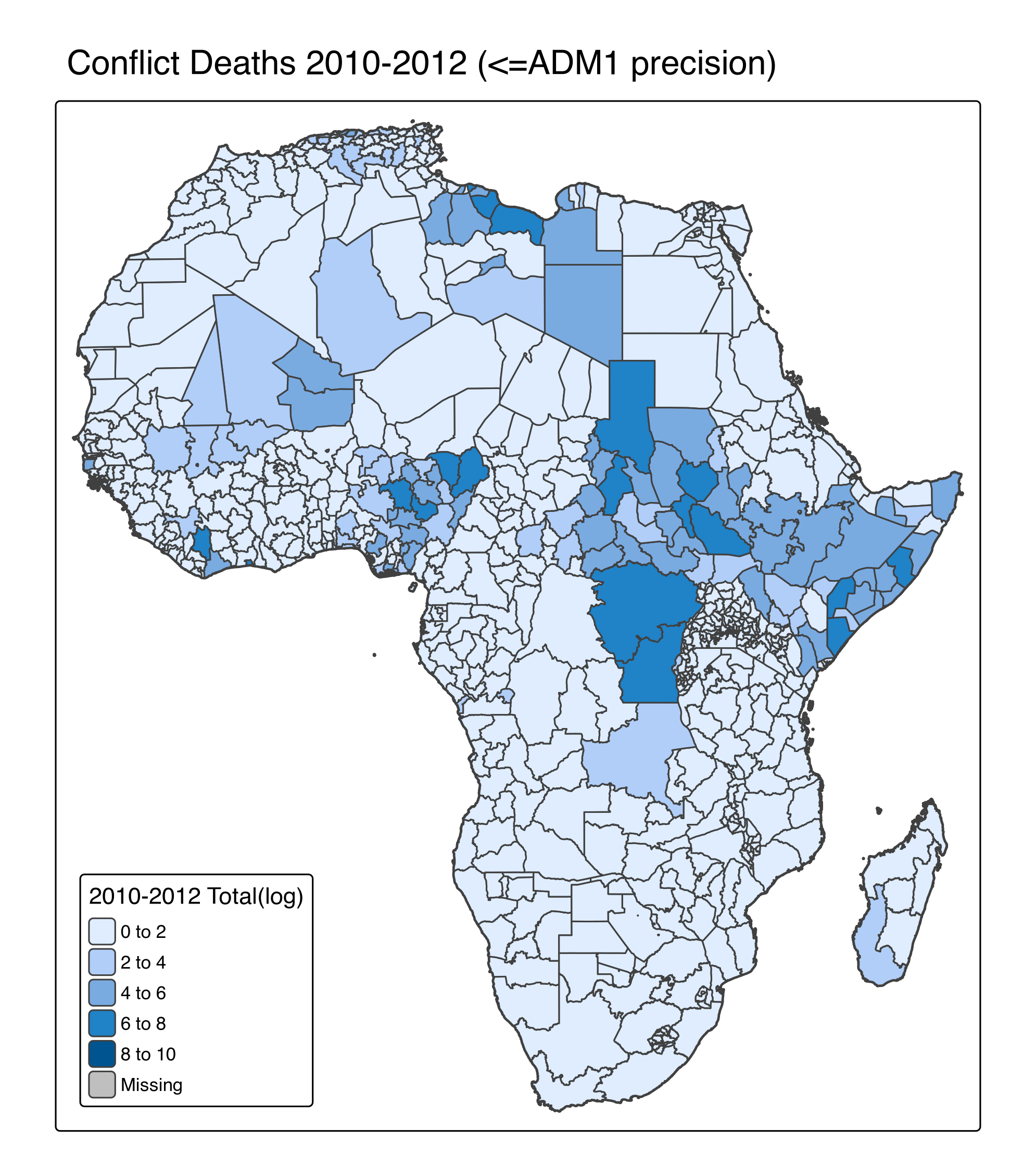}
\caption{Conflict deaths (2010-2012)}
\label{subfig:Conflict}
\end{subfigure}
\hfill
\begin{subfigure}[b]{0.45\textwidth}
\centering
\includegraphics[width=\textwidth,height=\textwidth,keepaspectratio=true]{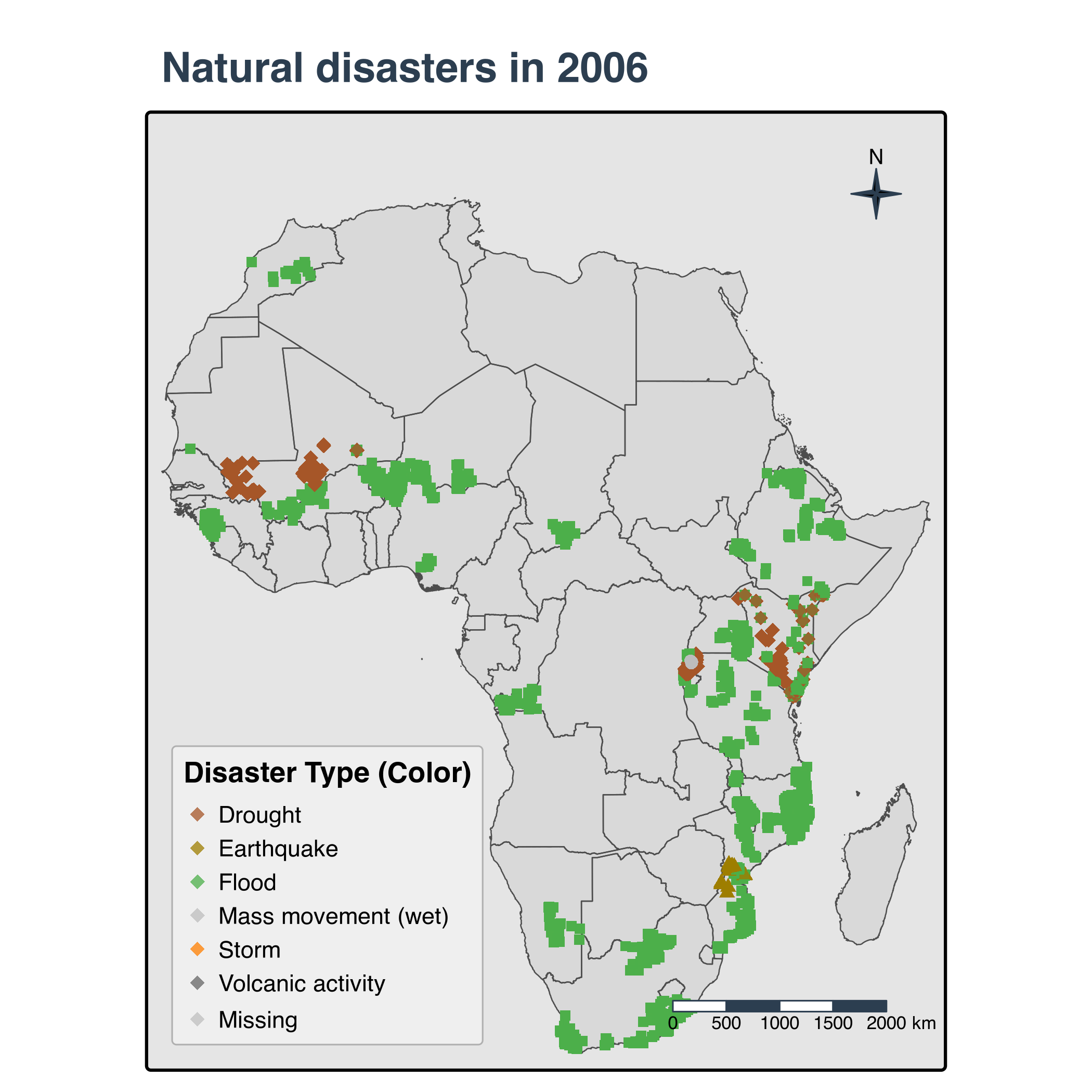}
\caption{Natural disasters (2006)}
\label{subfig:Disaster}
\end{subfigure}
\caption{Tabular covariate maps.}
\label{fig:FacetedCovariates}
\end{figure}


\clearpage

\section{Additional Empirical Results}

\begin{figure}[htb]
  \centering
  \includegraphics[width=1\linewidth]{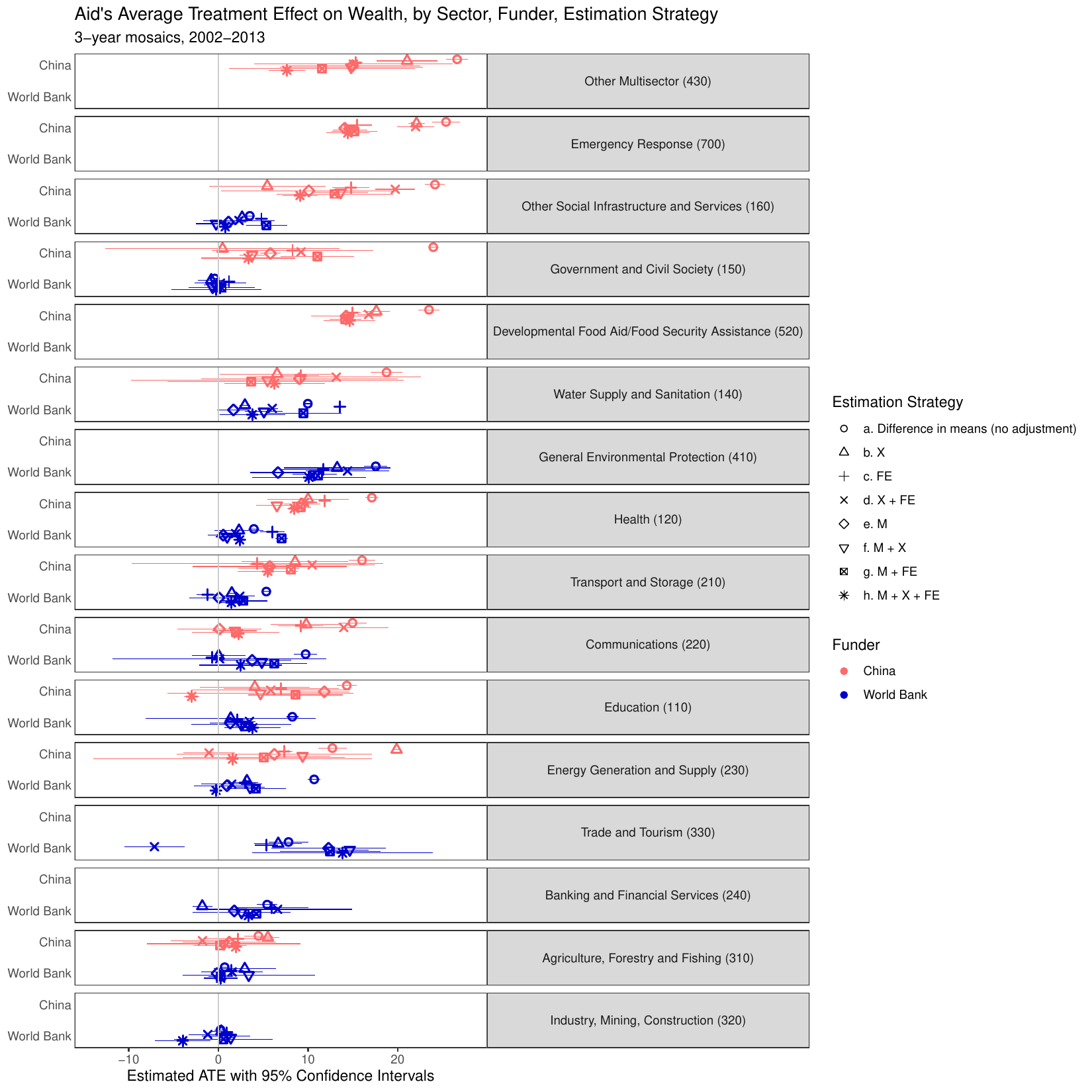}
  \caption{
ATE results, in full.
  }
\label{fig:ATEInFull}
\end{figure}

\begin{figure}[htb]
  \centering
  \includegraphics[width=1\linewidth]{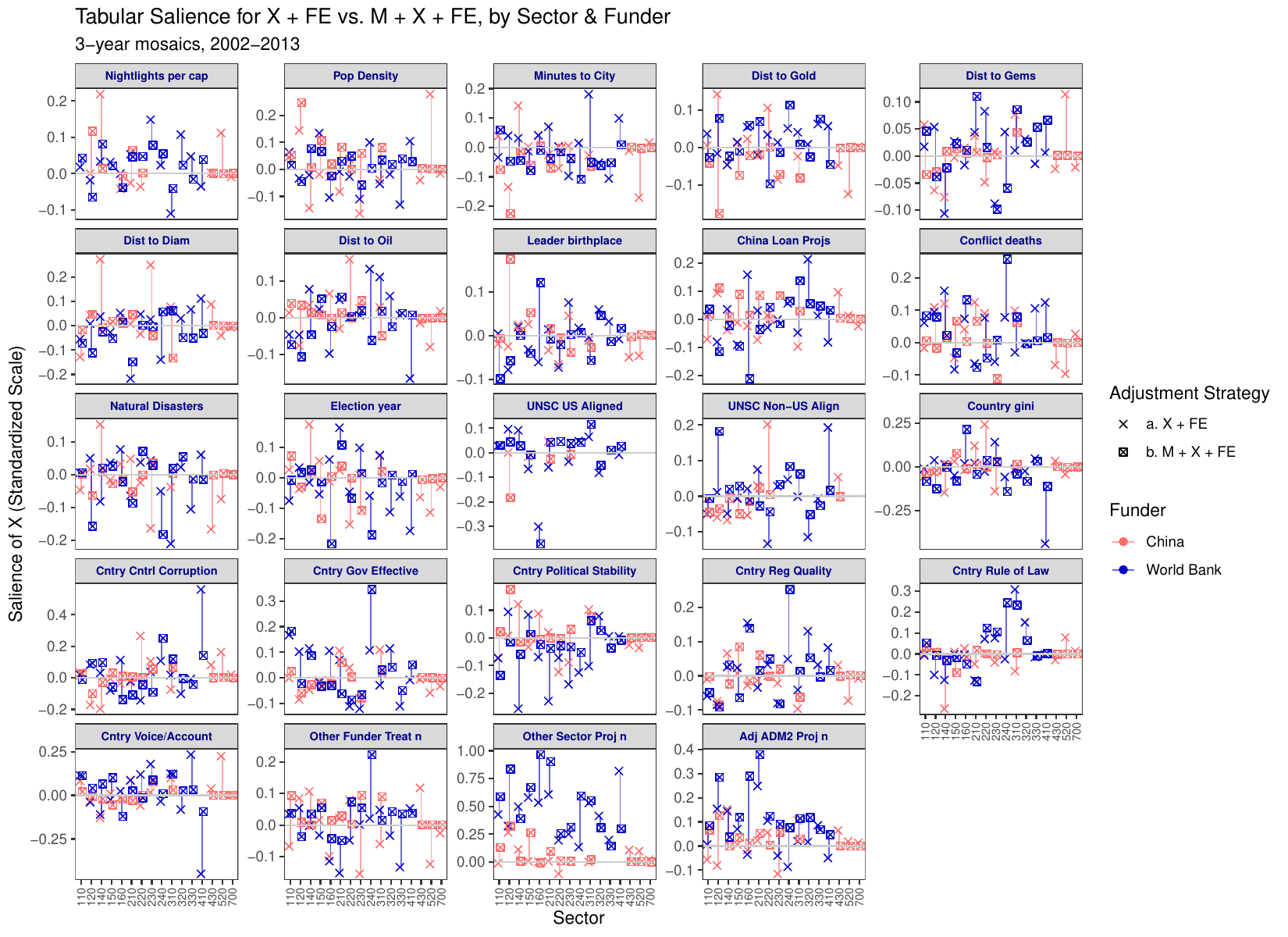}
  \caption{
 Salience of tabular variables to aid allocation across funders, sectors, and modeling strategy. This high-level overview indicates whether each variable had a positive, negative, or mixed effect on aid assignment for each sector ($y$-axis), and how the salience of each variable varied across sectors, funders, and modeling strategy.
  }
\label{fig:SALIENCETAB}
\end{figure}

\clearpage 

\section{Robustness Checks}

In this section, we present results from three robustness checks designed to assess the sensitivity of our main findings to specific methodological choices regarding covariate adjustment, outcome measurement resolution, and treatment definition.

\subsection{Sensitivity to Exclusion of Lagged Nighttime Lights}\label{s:FullNightlightRobustness}

\noindent \textbf{Motivation:} A potential concern in our analysis is the inclusion of lagged per-capita nightlights (NTL) in the covariate adjustment set. Because nightlights are highly persistent over time and are also used as features in the training of the machine-learning-derived International Wealth Index (IWI), their inclusion might introduce bias or over-control. To address this, we re-estimated our primary models after strictly excluding the lagged nightlights variable from the tabular covariate set.

\begin{figure}[htb]
  \centering
  \includegraphics[width=0.99\linewidth]{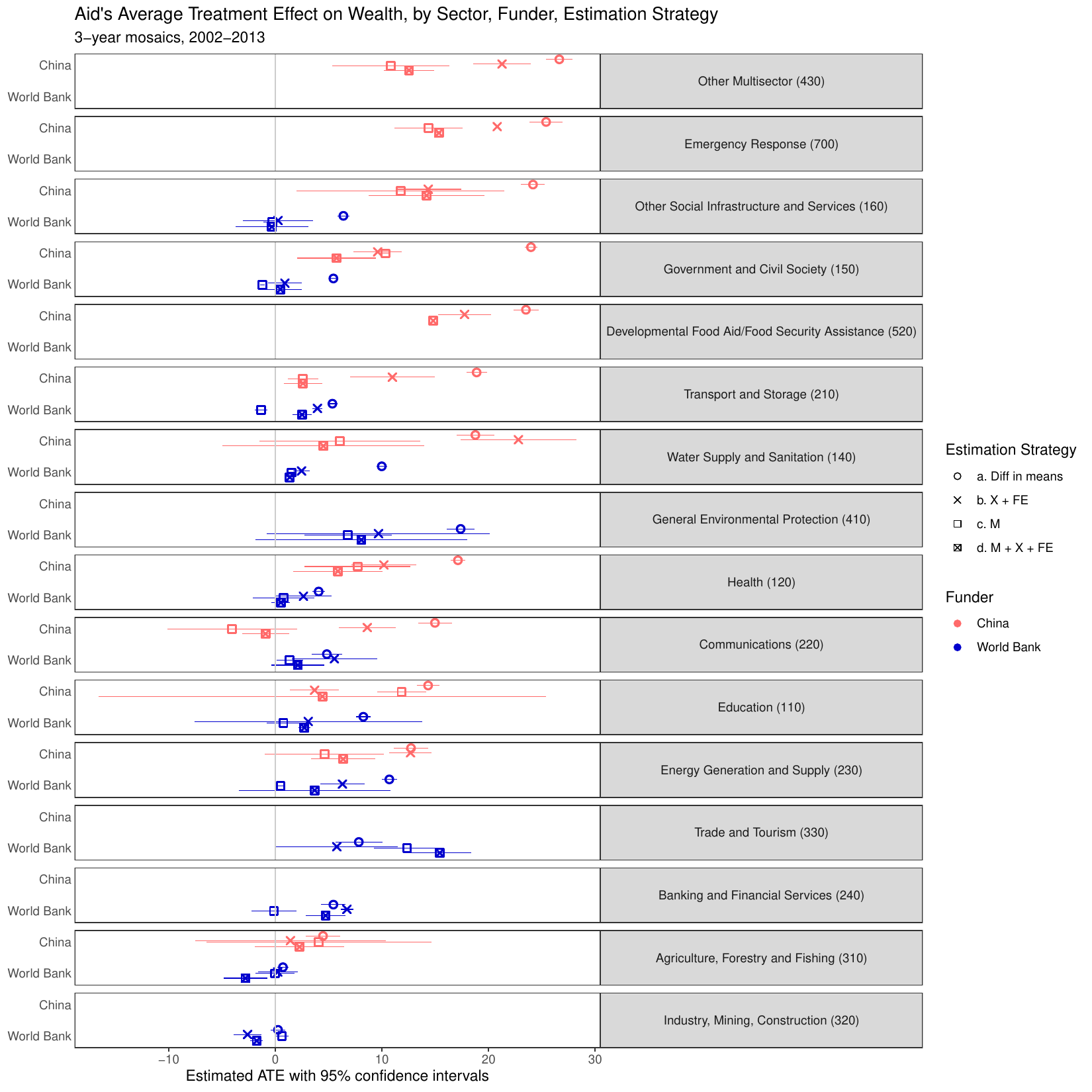}
  \caption{
    \textbf{Robustness (no nightlights):} ATEs by sector, funder, and modeling strategy, excluding lagged nightlights from the adjustment set.
  }
  \label{fig:ATE_NoNTL}
\end{figure}

\noindent \textbf{Results:} Figure \ref{fig:ATE_NoNTL} displays the Average Treatment Effects (ATEs) under this specification. The results are substantively similar to the main analysis, preserving the sectoral rankings and the general finding that Chinese projects are associated with larger wealth gains in the short term, at least using this identification strategy. This suggests our results are not driven by the mechanical reuse of nightlights data.

\subsection{Sensitivity to Neighborhood Size (Addressing Displacement)}\label{s:FullNeighborhoodSizeRobustness}

\noindent \textbf{Motivation:} The DHS cluster coordinates used to define our neighborhoods are randomly displaced by up to 2 km in urban areas and 5 km (or occasionally 10 km) in rural areas to protect respondent privacy. This displacement could misalign the treatment assignment with the outcome measurement (the 6.7 km $\times$ 6.7 km IWI raster). To address this, we re-estimated our models using a coarser outcome grid ($\approx$ 13.4 km $\times$ 13.4 km) created by averaging the four 6.7 km cells nearest to the cluster centroid.

\begin{figure}[htb]
  \centering
  \includegraphics[width=0.99\linewidth]{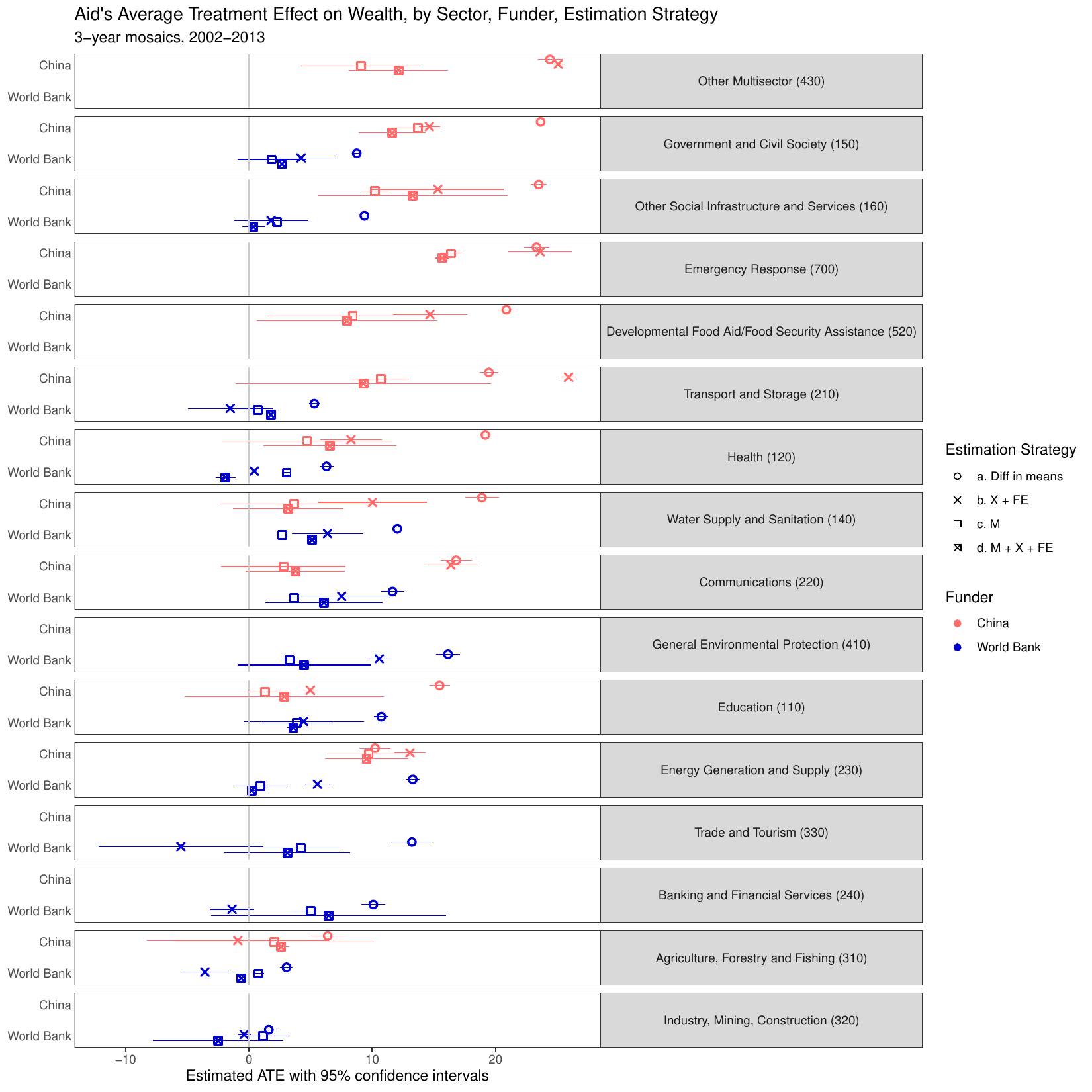}
  \caption{
    \textbf{Robustness (Coarser Outcome Raster):} ATEs by sector, funder, and modeling strategy using a $\sim$13.4 km outcome grid to account for DHS cluster displacement.
  }
  \label{fig:ATE_LargeBuff}
\end{figure}

\noindent \textbf{Results:} Figure \ref{fig:ATE_LargeBuff} presents the ATE estimates using this coarser outcome grid. While confidence intervals naturally widen slightly due to spatial aggregation, the point estimates and overall patterns remain robust. This indicates that our findings are not artifacts of spatial misalignment caused by DHS coordinate displacement, although we do not here address all forms of possible spillover. 

\subsection{Sensitivity to Strict Treatment Definition}\label{s:FullTreatmentDefineSensitivity}

\noindent \textbf{Motivation:} In our baseline analysis, we assigned treatment status based on project precision codes, including treating an entire ADM2 region if the project location was only known at the ADM2 level (Precision 3). To address concerns that this might dilute effects or introduce measurement error, we performed a "Strict Precision" robustness check. We restricted the treatment group to include only neighborhoods within 25 km of a precise project location (Precision 1 and 2), excluding those treated solely based on ADM2 boundaries.

\begin{figure}[htb]
  \centering
  \includegraphics[width=0.99\linewidth]{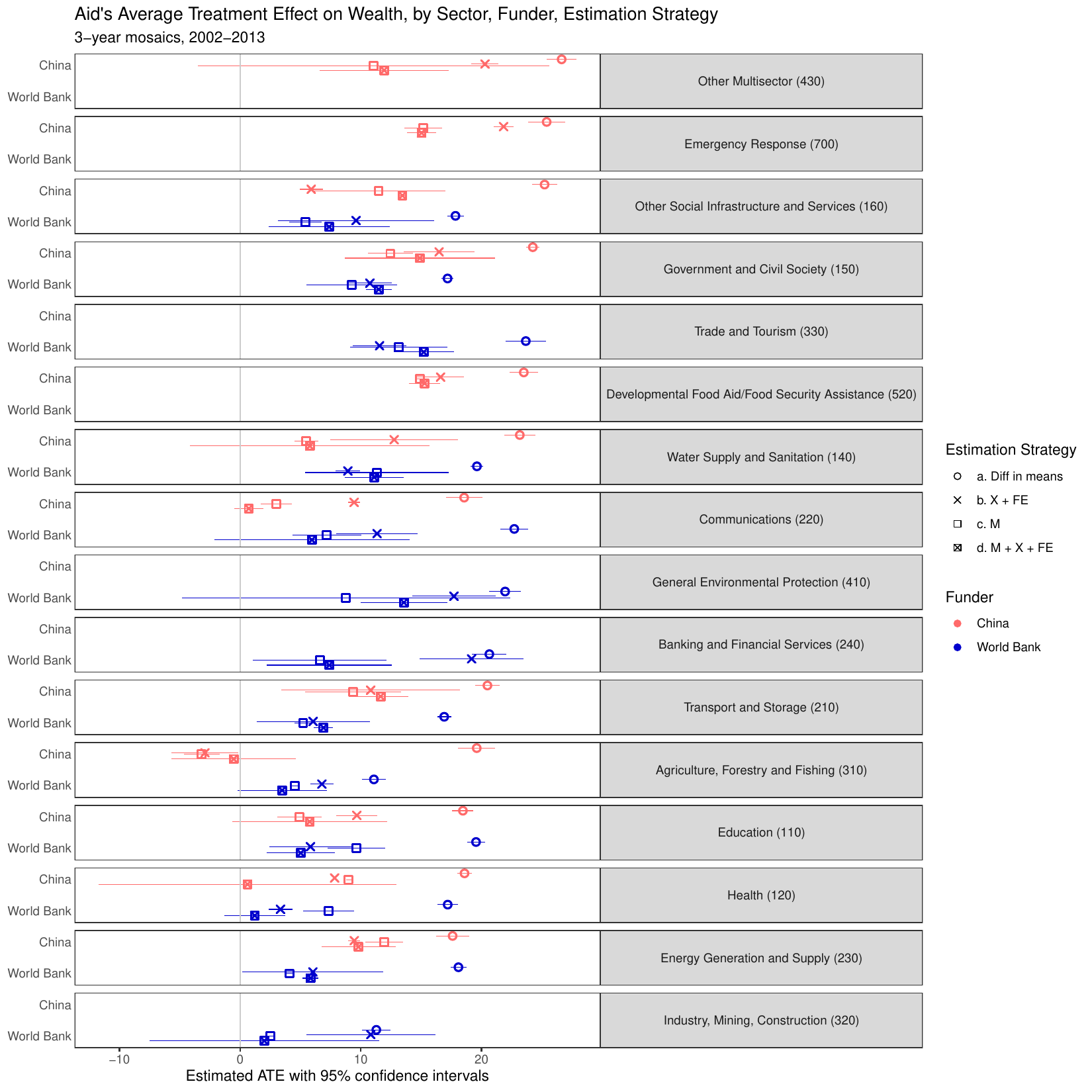}
  \caption{
    \textbf{Robustness (Strict Precision):} ATEs by sector, funder, and modeling strategy using a strict treatment definition (excluding ADM2-based assignment).
  }
  \label{fig:ATE_StrictPrec}
\end{figure}

\noindent \textbf{Results:} Figure \ref{fig:ATE_StrictPrec} displays the results under this stricter treatment regime. The estimated effects remain broadly consistent with the main analysis, although World Development estimates are notably higher in magnitude in this regime.

\end{document}